\documentclass[]{amsart}

\usepackage{manuscript}

\IfFileExists{./postpreamble-amspreprint.sty}{\RequirePackage[packages,changes,theorems]{postpreamble-amspreprint}}{}

\makeatletter
\@ifpackageloaded{changes}{
\definechangesauthor[name = {Hajg Jasa}, color = {TolVibrantCyan}]{HJ}
\definechangesauthor[name = {Ronny Bergmann}, color = {TolVibrantMagenta}]{RB}
\definechangesauthor[name = {Christian Kümmerle}, color = {TolVibrantTeal}]{CK}
\definechangesauthor[name = {Avanti Athreya}, color = {TolVibrantOrange}]{AA}
\definechangesauthor[name = {Zachary Lubberts}, color = {TolVibrantBlue}]{ZL}
}{}
\makeatother        

\makeatletter
\@ifpackageloaded{hyperref}{%
	\hypersetup{
		pdftitle = {Robust and Spectral Procrustes},
		pdfauthor = {Hajg Jasa, Ronny Bergmann, Christian Kümmerle, Avanti Athreya, Zachary Lubberts},
		pdfkeywords = {optimization on manifolds, spectral Procrustes},
		pdfcreator = {Created using the Scoop Template Engine version 1.4.2.}
	}
}{
	\pdfinfo{
		/Title (Spectral Procrustes)
		/Author (Hajg Jasa, Ronny Bergmann, Christian Kümmerle, Avanti Athreya, Zachary Lubberts)
		/Subject ()
		/Keywords (optimization on manifolds, spectral Procrustes)
		/Creator (Created using the Scoop Template Engine version 1.4.2.)
	}
}
\makeatother

\title[Procrustes Problems on Random Matrices]{Procrustes Problems on Random Matrices}

\author[H. Jasa]{Hajg Jasa\orcidlink{0009-0002-7917-0530}}
\address[H. Jasa]{Norwegian University of Science and Technology, Department of Mathematical Sciences, NO-7041 Trondheim, Norway}
\email{\detokenize{hajg.jasa@ntnu.no}}
\urladdr{https://www.ntnu.edu/employees/hajg.jasa}

\author[R. Bergmann]{Ronny Bergmann\orcidlink{0000-0001-8342-7218}}
\address[R. Bergmann]{Norwegian University of Science and Technology, Department of Mathematical Sciences, NO-7041 Trondheim, Norway}
\email{\detokenize{ronny.bergmannn@ntnu.no}}
\urladdr{https://www.ntnu.edu/employees/ronny.bergmann}

\author[C. Kümmerle]{Christian Kümmerle\orcidlink{0000-0001-9267-5379}}
\address[C. Kümmerle]{University of Central Florida, Department of Mathematics, Orlando, FL 32816, USA}
\email{\detokenize{kuemmerle@ucf.edu}}
\urladdr{https://ckuemmerle.com}

\author[A. Athreya]{Avanti Athreya}
\address[A. Athreya]{Department of Applied Mathematics and Statistics, Johns Hopkins University, Baltimore, MD 21218, USA}
\email{\detokenize{dathrey1@jhu.edu}}

\author[Z. Lubberts]{Zachary Lubberts}
\address[Z. Lubberts]{Department of Statistics, University of Virginia, Charlottesville, VA 22904, USA}
\email{\detokenize{zlubberts@virginia.edu}}
\urladdr{https://sites.google.com/view/zachary-lubberts/home}

\date{\today}

\dedicatory{}

\begin{document}

\begin{abstract}
Meaningful comparison between sets of observations often necessitates alignment or registration between them, and the resulting optimization problems range in complexity from those admitting simple closed-form solutions to those requiring advanced and novel techniques. We compare different Procrustes problems in which we align two sets of points after various perturbations by minimizing the norm of the difference between one matrix and an orthogonal transformation of the other. The minimization problem depends significantly on the choice of matrix norm; we highlight recent developments in nonsmooth Riemannian optimization and characterize which choices of norm work best for each perturbation. We show that in several applications, from low-dimensional alignments to hypothesis testing for random networks, when Procrustes alignment with the spectral or robust norm is the appropriate choice, it is often feasible to replace the computationally more expensive spectral and robust minimizers with their closed-form Frobenius-norm counterpart. Our work reinforces the synergy between optimization, geometry, and statistics.
\end{abstract}


\makeatletter
\ltx@ifpackageloaded{hyperref}{%
\subjclass[2020]{\href{https://mathscinet.ams.org/msc/msc2020.html?t=49Q99}{49Q99}, \href{https://mathscinet.ams.org/msc/msc2020.html?t=62-08}{62-08}, \href{https://mathscinet.ams.org/msc/msc2020.html?t=62F40}{62F40},
\href{https://mathscinet.ams.org/msc/msc2020.html?t=62-04}{62-04}}
}{%
}

\makeatother

\maketitle
%
%
\section{Introduction}
\label{section:introduction}

Alignment problems, in which comparing observations requires determining an orientation or correspondence across them, are important in multiple disciplines: image registration in computer vision \cite{xiong2010critical,Jubran-2021Provably}, graph matching in network analysis \cite{li2019graph,lyzinski16:_infoGM}, organoid correspondence in neuroscience \cite{chen2023discovering}, and distance computations in astronomy \cite{Wahba:1965, Markley:2000Quaternion}.  
Depending on the set of transformations to consider, the goal of these registration or Procrustes problems is to align two sets of points with a known correspondence in order to minimize a distance. This is a rich topic of interest optimization \cite{farrell1966least,de2013solution}, and the choice of distance, in particular the choice of matrix norm, has implications for the optimization techniques required. For instance, the classical Procrustes problem seeks to minimize the Frobenius norm between two given matrices, one of which has undergone an orthogonal transformation. In this case, the minimizing orthogonal transformation has a closed-form solution \cite{cape2019two}. If the Frobenius norm is replaced by the spectral norm, however, the problem changes dramatically; a computationally feasible approach to this nonsmooth optimization is a recent development \cite{BergmannHerzogJasa:2024}. The impact of the choice of matrix norm, both in terms of a specific data analysis task and in terms of computational complexity, warrants closer examination of Procrustes problems for random or noisily observed matrices.

While the Frobenius norm {\em per se} results in a nonsmooth optimization itself, it shares its minimizer with its square, which is smooth and can be used to obtain a closed form solution. Minimizing the spectral norm results in a nonsmooth one without a closed form solution.
We also consider what is sometimes called the \emph{robust norm} 
(see \eqref{eq:RobustProcrustes} below) \cite{Eldar-2009Robust,Bustos-2017Guaranteed}, which inherently penalizes outlier points, and which has many more discontinuities in its derivative. All three problems can be phrased as optimization problems either with constraints and smoothing as in~\cite{Nesterov:2006} or on the Riemannian manifold of orthogonal matrices.
In recent years, algorithms for smooth optimization have been investigated and generalized to solve optimization problems on Riemannian manifolds~\cite{AbsilMahonySepulchre:2008,Boumal:2023}.
For nonsmooth optimization problems on Riemannian manifolds such as spectral and robust minimization, algorithms can typically be split into three categories. If one has no derivative information, one can employ Riemannian variants of Nelder-Mead~\cite{Dreisigmeyer:2007}, particle swarm~\cite{BorckmansIshtevaAbsil:2010}, or mesh adaptive algorithms~\cite{Dreisigmeyer:2007}.
If one has first-order derivative information, one can use the subgradient method~\cite{FerreiraOliveira:1998}; the proximal point algorithm~\cite{FerreiraOliveira:2002}; or methods that collect subgradients, the so-called bundle methods \cite{BergmannHerzogJasa:2024,HoseiniMonjeziNobakhtianPouryayevali:2021}.
As an alternative, Riemannian smoothing methods are proposed in~\cite{PengWuHuDeng:2023, ZhangChenMa:2024}.
Beyond that, if the cost can be split into summands, splitting-based methods have been applied~\cite{Bacak:2014,BergmannHerzogSilvaLouzeiroTenbrinckVidalNunez:2021,BergmannJasaJohnPfeffer:2025, BergmannPerschSteidl:2016}; on embedded manifolds these can also be combined with smoothing~\cite{BeckRosset:2023} .

As a concrete example of the utility of Procrustes minimizations under various norms, consider the inference task of determining whether two random networks are statistically similar \cite{tang2017semiparametric}. If both random networks are instantiations of latent position random networks \cite{Hoff2002}, in which probabilities of connection are functions of matrices of low-dimensional features, a standard approach to such a hypothesis test is to spectrally embed the two observed adjacency matrices of the random networks into a lower-dimensional space via a truncated singular value decomposition, and compute a Procrustes minimization between these two singular value decompositions \cite{athreya2017statistical}. Under mild assumptions, the embeddings of the observed network adjacencies provide consistent estimates for the network's underlying features \cite{STFP-2011}, and the Procrustes minimization considers how different these features are after an alignment. Namely, if $\hat{X}, \hat{Y} \in \mathbb{R}^{n \times d}$ denote the embeddings of the observed adjacency matrices for the first and second networks, respectively, one can compute
\begin{equation}
    \label{eq:Frob_norm_test_stat}
    T_F
    \coloneqq
    \min_{W \in \mathbb{R}^{d \times d}, W^{\top} W=I}
    \lVert 
        \hat{X} - \hat{Y}W
    \rVert_F
    .
\end{equation}
If this quantity is large, we reject the null hypothesis that the two networks have the same underlying features. The Procrustes minimization over orthogonal matrices resolves the inherent non-identifiability of the singular value decomposition and is necessary for comparison between the two computed embeddings.
As previously mentioned, when using the Frobenius norm, the optimal solution for the minimizing orthogonal transformation can be exactly computed. However, the Frobenius norm provides but one measure of the dissimilarity between corresponding eigenspaces. For instance, the Frobenius norm cannot distinguish between the case in which most eigendirections of the underlying network features are perfectly aligned and only a few are not, and the cases in which all eigendirections have moderate errors. To address the first, one can consider the spectral norm, as in \cite{Athreya:2025EuclideanMirrors}. 
The spectral norm allows for more nuanced comparisons between matrices of eigenvectors and can yield useful theoretical results, but the lack of a closed-form solution for the minimizing orthogonal transformation is an impediment to its use. As a practical matter, the closed-form Frobenius-norm minimizer can be substituted for the spectral-norm-minimizing orthogonal transformation (see \cite{chen2023discovering, chen2024euclidean}). The replacement of the Frobenius-norm minimizer in the spectral-norm objective gives a measure of alignment between two matrices, and as we show, can be an effective approximation. Nevertheless, the consequences of such an approximation are delicate, especially in hypothesis testing. One motivation for this work is to better understand these consequences.

 Suppose the spectral norm is used in Eq. \eqref{eq:Frob_norm_test_stat}; denote the resulting test statistic by $T_S$. Again, if the null hypothesis is equality of underlying network features, we reject the null if the observed value of $T_S$ is large. An approximation of $T_S$ with the Frobenius norm minimizer $W_F^{*}$ incurs both a model-based variability, inherent in the test statistic with the (correct) spectral-norm minimizing orthogonal matrix $W^*_S$, {\em and} a numerical variability that arises from the gap in the objective function when using $W^*_F$ as opposed to $W^*_S$. Further, the behavior of the Frobenius objective function compared to its spectral counterpart also plays a role; even with these two sources of variation in the approximation, using the spectral norm with Frobenius-norm minimizer can produce a test statistic that exhibits empirically greater power than using the spectral norm with its true minimizer.

 Typically, the allowable Type I Error probability, or significance level, of a hypothesis test governs the specific set of values of the test statistic for which we reject the null hypothesis. In our case, we reject the null for values of the test statistic that are sufficiently large; that is, values of the test statistic that exceed some critical value. Given a level $\alpha \in (0,1)$, the critical value $c_{\alpha}$ satisfies the requirement that the probability of $T_S$ exceeding $c_{\alpha}$ under the null is at most $\alpha$. Since the Procrustes problem is a minimization problem, the  statistic $\hat{T}_S=\|\hat{X}-\hat{Y}W_F^*\|_2$ always serves as an upper bound for $T_S=\|\hat{X}-\hat{Y}W_S^*\|_2$, where $W_F^*,W_S^*$ minimize the optimization problem of Eq.~\eqref{eq:Frob_norm_test_stat} for the Frobenius and spectral norms, respectively. Let $c_{\alpha}$ denote the true critical value for the test statistic $T_S$. Since $\hat{T}_S \geq T_S$, we see that if $T_S>c_{\alpha}$, then $\hat{T}_S \geq T_S >c_{\alpha}$, and both the approximate and correct test statistic yield the same test outcome.  However, if $T_S<c_{\alpha}$, it is entirely possible that $\hat{T}_S$ exceeds $c_{\alpha}$, leading to a divergence in test results between the true and approximated test statistic, and making the appropriate choice of critical value more difficult to determine.

Of course, the theoretical critical value $c_{\alpha}$ is often unknown. For hypothesis testing on networks, this value can be estimated via a parametric bootstrap \cite{efron1994introduction, levin2025bootstrapping}, in which observed values of networks are used to generate subsequent network realizations, from which empirical distributions of test statistics can be computed. Let $\hat{c}_{\alpha}$ denote such an estimate. If we use a test statistic with an approximate minimizer, the $(1-\alpha)$-quantile of the distribution will be larger than the true critical value; that is, $c_{\alpha}< \hat{c}_{\alpha}$. This gives a more conservative test, but at the cost of a decrease in power.
Depending on the size of the gap between $\hat{T}$ and $T$, the test statistic with an approximate minimizer $\hat{T}$ exceeding the estimated critical value $\hat{c}_{\alpha}$ may not correspond to the test statistic with the correct minimizer $T$ exceeding the correct critical value $c_{\alpha}$. Such information loss is undesirable.

We focus on the power of different Procrustes-based statistics in hypothesis testing for network comparison over a range of alternative hypotheses. The Procrustes minimizations we consider are ubiquitous in network statistics; the improved optimization methods we highlight allow statisticians to solve them more precisely. Further, as our simulations demonstrate, heuristics and approximate solutions, such as the substitution of the Frobenius-norm minimizer into alignment problems with different matrix norms, can work surprisingly well for inference tasks. This lends credence to these computationally inexpensive shortcuts, and suggests several downstream research questions on trade-offs between computational and statistical error, appropriate choice of matrix norm for a particular inference task, and the interplay between an objective function and the distribution of the resulting test statistic. A more nuanced understanding of the challenges and implications of Procrustes problems for random matrices, for optimization and statistics, is a novel contribution of this work. 

We organize the paper as follows. In Section \ref{section:procrustes}, we introduce general terminology and background for Procrustes problems. In Section \ref{sec:optimization_on_manifolds}, we highlight notions from differential geometry relevant to optimization on manifolds. In Section \ref{section:statistical}, we discuss network models and hypothesis testing for random networks. In Section \ref{sec:Duck_example}, we provide a visualizable and concrete example of perturbations in two dimensions and investigate the consequence of Procrustes alignments with different norms. In Section \ref{sec:hypothesis_testing}, we showcase the impact of Procustres-based test statistics on statistical power for two-sample hypothesis tests on networks. In Section \ref{sec:discussion}, we identify directions for ongoing and future research.

\section{Procrustes Problems}%
\label{section:procrustes}

The Procrustes problem is to align or match the columns of two matrices $A,B \in \mathbb R^{d\times n}$ after an appropriate orthogonal transformation, namely after multiplying one of the matrices by a matrix $W\in \mathbb R^{d\times d}$ satisfying $W^{\top}W = I$, where $I \in \mathbb R^{d\times d}$ denotes the identity matrix. See \cite{GowerDijksterhuis-2004} for a comprehensive overview. 
Classically, one considers the \emph{Orthogonal Procrustes} problem, which is defined with respect to the Frobenius norm:
\begin{equation}\label{eq:OrthogonalProcrustes}
    W_{\mathrm{F}}^* \in \argmin_{W\in \mathbb R^{d\times d}, W^{\top}W = I} 
    f_{\mathrm{F}}(W),
    \qquad f_{\mathrm{F}}(W) \coloneqq \lVert A - WB \rVert_{\mathrm{F}}
    .
\end{equation}

Formally, both $f_{\mathrm{F}}$ and $W_{\mathrm{F}}^*$ depend on the matrices $A,B$. We omit those, however, since they are clear from context in the following. The closed-form solution of this problem is given by the singular value decomposition (SVD) of $B A^\top = U\Lambda V^\top$.
In this case, $W_{\mathrm{F}}^* = VU^\top $. Since the Frobenius norm of a matrix is the square root of the sum of the squares of all of its entries, it 
is sensitive to errors in every matrix entry. From a statistical perspective, this makes it a reasonable choice for measuring and detecting small, diffuse errors.

However, when errors behave differently, for example in the case of a low-rank perturbation or a few large outliers, then other norms may be preferred. For example,~\cite{Verboon-1992Resistant} considered Huber-norm based formulations, and~\cite{Andreella-2023Procrustes, Gower-1975Generalized} generalized the task to include translations and scaling.
Motivated by our statistical network applications, in this work, we consider different choices of matrix norm for measuring alignment.

The \emph{Spectral Procrustes} problem \cite{Athreya:2025EuclideanMirrors} uses the spectral norm to compare $A$ to the modified version $WB$ of $B$, resulting in the problem:
\begin{equation}\label{eq:SpectralProcrustes}
    W_{\mathrm{S}}^* \in \argmin_{W\in \mathbb R^{d\times d}, W^{\top}W = I} 
    f_{\mathrm{S}}(W),\qquad f_{\mathrm{S}}(W) \coloneqq 
    \lVert A - WB \rVert_{\mathrm{S}}
    ,
\end{equation}
where $\lVert A\rVert_{\mathrm{S}}\coloneqq \lVert A \rVert_2$ denotes the spectral norm of the matrix $A$, \ie its largest singular value. We again omit $A,B$ in the notation, since the matrices are clear from context.

Intuitively, this is more robust to noise because the spectral norm only depends on the largest singular value of a matrix, while the closed-form solution of the orthogonal Procrustes problem depends on all of the singular values. On the other hand, this problem does not yield a closed-form solution, and it may not be differentiable.  This can be seen most easily by considering the formulation of the spectral norm as 
$$
\|A\|_2 = \max_{\substack{x\in \mathbb R^d,\,y\in \mathbb R^n\\ \|x\|_2=\|y\|_2=1}} x^\top A y.
$$
Since this is a maximum of linear functions of $A$, it is clearly convex (on $\mathbb R^{n\times d}$), and when defined, its gradient is given by $x_1 y_1^\top$, for $x_1,y_1$ a pair of vectors achieving this maximum. However, when multiple linear functions yield the maximum value (which occurs only when the largest singular value of $A$ has multiplicity exceeding 1), the norm will not be differentiable.

Finally, we consider as a third norm, the outlier-robust norm
\begin{equation*}
    \lVert A \rVert_{\mathrm{R}} = \sum_{k=1}^n \Bigl(
        \sum_{j=1}^d a_{jk}^2
    \Bigr)^\frac{1}{2} = \sum_{k=1}^n \lVert A_{:,k} \rVert_2.
\end{equation*}
for a matrix $A = (a_{jk})_{j=1,k=1}^{d,n} \in \mathbb R^{d\times n}$, which is called the mixed $\ell_2/\ell_1$ norm or $\ell_{2,1}$-norm in the compressed sensing literature \cite{Kowalski-2009Sparse,Eldar-2009Robust,Bach-2012Optimization,Elhamifar-2012Block}.\footnote{We note that this norm does \emph{not} coincide with the vector-norm induced matrix norm $\|A\|_{2 \to 1}= \sup_{x:\|x\|_1\leq 1} \|Ax||_1$.} We define the associated so-called \emph{Robust Procrustes} problem as 
\begin{equation}\label{eq:RobustProcrustes}
    W_{\mathrm{R}}^* \in \argmin_{W\in \mathbb R^{d\times d}, W^{\top}W = I} f_{\mathrm{R}}(W),
    \qquad f_{\mathrm{R}}(W) \coloneqq \lVert A - WB \rVert_{\mathrm{R}},
\end{equation}
The robust norm $\lVert\cdot\rVert_{\mathrm{R}}$ takes the standard Euclidean norm $\lVert A_{:,k} \rVert_2$ of every column $A_{:,k} = \begin{pmatrix} a_{1k}, & \ldots, & a_{d k} \end{pmatrix}$ of $A$, then applies the one-norm to the resulting vector of norms. Similar to the $\ell_1$-norm used in sparse and robust regression \cite{Hastie-2015Statistical}, the minimization of $f_{\mathrm{R}}(\cdot)$ encourages the alignment of \emph{most} of the columns of $A$ and $B$, but allows for a mismatch of \emph{few} columns. 
The challenge of \eqref{eq:RobustProcrustes} is that it does not have a closed-form solution,  unlike \eqref{eq:OrthogonalProcrustes}, and is a nonsmooth optimization problem with non-convex domain $\{W\in \mathbb R^{d\times d}, W^{\top}W = I\}$ as neither $\lVert\cdot\rVert_{\mathrm{R}}$ nor its square are differentiable. 
A related robust variant of the Procrustes problem is considered in \cite{Verboon-1992Resistant}, and more recently, more general Procrustes-like problem formulations with outlier-robust penalizations have gained attention for solving outlier-robust point registration problems in computer vision \cite{Yang-2019Quaternion,Huang2024Scalable}.

\section{Optimization on Manifolds}\label{sec:optimization_on_manifolds}
In order to provide algorithms that enable solving each of the generalized Procrustes problems \eqref{eq:OrthogonalProcrustes},  \eqref{eq:SpectralProcrustes} and \eqref{eq:RobustProcrustes}, we consider the framework of optimization on Riemannian manifolds \cite{AbsilMahonySepulchre:2008,Boumal:2023}.
A $d$-dimensional Riemannian manifold is a set $\mathcal M$ that is locally homeomorphic to $\mathbb R^d$. While this is usually written with charts, see, \eg, \cite{DoCarmo:1992}, in algorithms it is often preferable to work chart-free.
A manifold is equipped with tangent spaces: At a point $p\in \mathcal M$ consider all curves $c\colon (-\varepsilon,\varepsilon) \to \mathcal M$ for some small $\varepsilon > 0$ such that $c(0)=p$ and declare two such curves equivalent if they have the same derivative at $0$. The set of such equivalence classes is called the \emph{tangent space} $\tangentSpace{p}$ at $p$. This space is a vector space. If we equip these tangent spaces with inner products $g_p$ such that the mapping $p \mapsto g_p$ is smooth, we obtain a Riemannian manifold.
A prominent example is the unit sphere $\mathbb S^2 = \{ p \in \mathbb R^3, \lVert p \rVert = 1\}$ where the tangent spaces are the tangent planes in the embedding $\mathbb R^3$ and restricting the Euclidean inner product to these tangent spaces yields a Riemannian manifold.

We need two further concepts on manifolds. Given a point $p\in$ $\cM$ and a tangent vector $X \in \tangentSpace{p}$, we consider a curve $c\colon I \to \mathcal M$, $[0,1] \subset I$, such that $c(0) = p$ and $c'(0) = X$. Then, the map $R_p \colon \tangentSpace{p} \to \mathcal M$ defined by $R_p(X) = c(1)$ is called a \emph{retraction}. It allows us to “move in the direction $X$” along the manifold. One also needs to ensure that the \emph{differential}, which is a linear map $\mathrm{d}_X R_p \colon \tangentSpace{p} \to \tangentSpace{R_p(X)}$, of $R_p$ at the point $0 \in \tangentSpace{p}$ is the identity, \ie, $d_{0}R_p[Y] = Y$ for all $Y \in \tangentSpace{p}$. A special retraction is the \emph{exponential map} $\exp_p\colon \tangentSpace{p} \to \cM$, when the curves are chosen to be geodesics~\cite[Def.~5.38]{Boumal:2023}.
A \emph{geodesic} is the generalization of a straight line to Riemannian manifolds, characterized by having zero acceleration. 
Locally around the origin $0 \in \tangentSpace{p}$, the exponential map is a diffeomorphism. We denote by $B_p(r)$ the open ball of radius $r$ around the origin $0 \in \tangentSpace{p}$. The largest radius $r$ such that the exponential map is a diffeomorphism between $B_p(r)$ and $\exp_p(B_p(r))$ is called the \emph{injectivity radius at $p \in \cM$}. 
The \emph{injectivity radius of $\cM$} is the infimum of the injectivity radius at $p$, over all $p \in \cM$.
For example, on the sphere $\mathbb S^2$, the injectivity radius is $\pi$.
Finally, a \emph{vector transport} $\vectorTransport{p}{q}\colon T_p\mathcal M \to T_q\mathcal M$ \cite[Def.~10.62]{Boumal:2023} is a similar way to “move” tangent vectors from the tangent space $\tangentSpace{p}$ to $\tangentSpace{q}$.

The following manifolds are of interest to us. The \emph{orthogonal group} $\mathrm{O}(d)$ is the set $O(d) \coloneqq \{ p \in \mathbb R^{d\times d}, p^{\top}p = I\}$ equipped with the standard matrix product. Notice that the problems in Eqs.~\eqref{eq:OrthogonalProcrustes}, \eqref{eq:SpectralProcrustes}, and \eqref{eq:RobustProcrustes} can be seen as optimization tasks over $\mathrm{O}(d)$.
This manifold actually consists of two disjoint sets, those of determinant $\det(p)=1$ and those of determinant $\det(p)=-1$. 
A special element is the identity matrix $I_d \in \mathbb R^{d \times d}$ within the first case. From the idea of charts, or equivalently following curves, all points in $O(d)$ with determinant 1 can be reached from $I_n$. This component of $\mathrm{O}(d)$ is also called the \emph{special orthogonal group} $\mathrm{SO}(d)$. This can be interpreted as the set of all $d$-by-$d$ rotation matrices. Both a manifold of unit determinant matrices and the rotation matrices are available in the Julia package~\texttt{Manifolds.jl}~\cite{AxenBaranBergmannRzecki:2023}. 

As mentioned above, a challenge with both spectral and robust Procrustes is that they are nonsmooth optimization problems. That is, the objective (or cost) function $f\colon\mathcal M \to \mathbb R$ to be optimized is not necessarily differentiable. Therefore, we use the lower triangle mesh-adaptive direct search (LTMADS) method \cite{Dreisigmeyer:2007} to address this issue.
MADS methods work in two steps: A \emph{search} and a \emph{poll step}. In the \emph{search step}, they employ the tangent space $\tangentSpace{p^{(k)}}$ at the current iterate $p^{(k)}$ and a mesh size $\Delta_k$ to build a mesh. On a manifold, this is done in coefficients with respect to a tangent space basis. The mesh directions are chosen with a binary vector $b_l$ and a lower triangular matrix $L$, hence the name LTMADS; for every direction generalized, its opposite direction is also added to the set of directions. The best of these directions is chosen in the \emph{poll step}. For each of the directions $d_j \in \tangentSpace{p^{(n)}}, j=1,\ldots,2d$ we evaluate the function at points in the manifold using the retraction $R_{p^{(k)}}$, obtaining $f(R_{p^{(k)}}(\Delta_kd_j))$, and select the first that improves the function value, \ie, where this value is less than $f(p^{(k)})$. Then we set the next iterate to $R_{p^{(k)}}(\Delta_kd_j)$ for the index $j$ that improved the cost.
If we find such a direction, the search step explores the direction found in the last poll by trying a step length of $4\Delta_k$ and using $R_{p^{(k)}}(4\Delta_kd_j)$ and keeps this as a new iterate if it further decreases the cost. 

If we find a new iterate in either the poll or the search step, we transport the tangent space basis to the new iterate $p^{(k+1)}$.
Finally, we update the mesh size $\Delta_k$. We set $\Delta_{k+1} = \frac{1}{4}\Delta_k$ if neither the poll nor the search find a point to improve the cost. If a better point is found, we increase the step size $\Delta_{k+1} = 4\Delta_k$ if $\Delta_k < \frac{1}{4}$ and keep the same mesh size otherwise.
One can also scale the whole mesh by a global factor $s$, but on manifolds one has to be careful to choose it less than the injectivity radius at the iterate.
As a stopping criterion, we use a lower threshold on the mesh size, \eg machine precision. This algorithm is available in \texttt{Manopt.jl} \cite{Bergmann:2022}\footnote{See \url{https://manoptjl.org/stable/solvers/mesh_adaptive_direct_search/}.}
and can be used with any manifold defined using \texttt{ManifoldsBase.jl}, especially all defined in \texttt{Manifolds.jl}.
An important concept in Riemannian optimization is that of \emph{geodesic convexity}.
A function $f \colon \cM \to \bbR$ is said to be geodesically convex if its composition $f \circ \gamma$ with any geodesic $\gamma$ of $\cM$ is convex in the usual sense.
Since the functions we consider, such as the spectral norm and the robust norm in Equations \eqref{eq:SpectralProcrustes} and \eqref{eq:RobustProcrustes}, may not be geodesically convex in general (see Figure~\ref{fig:costTildeD}), there is no guarantee of reaching a global minimizer a priori.
However, what we do obtain is a stationary point of the objective function. Applying this to Procrustes problems, this means that we have an optimization framework to find stationary points of \eqref{eq:SpectralProcrustes} and \eqref{eq:RobustProcrustes}.
It would be helpful to understand how benign the non-convexity introduced by the orthogonal group $O(d)$ is for the objectives at hand, which remains a topic for future research. 

\section{Statistical Background}
\label{section:statistical}

In this section, we consider the semiparametric problem of comparing the distributions of two random graphs \cite{tang2017semiparametric}.
For background, an undirected \emph{graph} is a pair $(V,E)$ where $V$ is a set of \emph{vertices} and $E$ a set of \emph{edges}, each of which consists of a pair of distinct vertices, namely $e=\{u,v\}\subseteq V$. \emph{Random dot product graphs} (RDPGs) \cite{young2007random} describe a probability distribution over such graphs by associating to each vertex $v$ a (generally unobserved) \emph{latent position} $x_v\in \RR^d$. The probability of observing an edge between two vertices $u$ and $v$ in the graph is given by $P_{uv}=\langle x_u,x_v\rangle$ for $u,v\in V$, and all edges arise independently. Because of its easy interpretation and flexibility, this is a workhorse model for random networks \cite{athreya2017statistical}.

If we consider every possible pair of edges $\{u,v\}$, $u,v \in V$, we can represent a graph as a symmetric, binary \emph{adjacency matrix} with $A_{uv}=1$ when $\{u,v\}\in E$, and $A_{uv}=0$ otherwise. Assuming that we have an RDPG, the mean of the adjacency matrix can be defined entrywise as
$$P_{uv}=\EE[A_{uv}]=\PP[A_{uv}=1]=\PP[\{u,v\}\in E]=\langle x_u,x_v\rangle=(XX^{\top})_{uv},$$ where $X\in \RR^{n\times d}$ collects the latent positions $x_u^{\top}$ as its rows. Since the mean probability matrix $P$ may be factored as $XX^{\top}$, its rank can be at most $d$.
Let $P=USU^{\top}$, where $S\in \RR^{d\times d}$ is a diagonal matrix with positive diagonal entries. Then $US^{1/2}=XW$ for some orthogonal matrix $W$ \cite{HJ85_matrix}. This non-identifiability up to an orthogonal matrix is unavoidable, since
\begin{equation*}
P_{uv}=\langle x_u, x_v\rangle = \langle Wx_u, Wx_v\rangle.
\end{equation*}
The realized or observed adjacency matrix of connections $A$ for this network provides noisy information about $P$, but even given $P$, the orthogonal non-identifiability allows us to recover the matrix of latent positions $X$ only up to an orthogonal transformation $W$. However, we may use the scaled eigenvectors of the observed adjacency matrix $A=\hat{U}\hat{S}\hat{U}^\top$ to obtain an estimate $\hat{X}=\hat{U}\hat{S}^{1/2}$ which is close to the matrix $XW$.
The estimate $\hat{X}$ is called the \emph{adjacency spectral embedding (ASE)} \cite{STFP-2011} of $A$.

Suppose we observe two vertex-aligned random dot product graphs $G_1$ and $G_2$. A natural graph inference question is to determine whether or not the two networks have the same underlying matrices of latent positions, up to an orthogonal transformation (if they do, the connection probabilities between vertices are the same for both networks). This is the \emph{semiparametric} test of hypothesis for the equality of RDPG models \cite{tang2017semiparametric}. When $X_1=X_2 W$, we have the exact same matrix of connection probabilities for every edge, and since edges arise independently in the RDPG model, this is sufficient for the network distributions to be the same. As such, an intuitive test statistic for this hypothesis takes the form 
\begin{equation} 
\label{eq:teststat}
T_C=\min_{W\in \mathbb R^{d\times d}, W^{\top}W = I} \|X_1^{\top} - W X_2^{\top} \|_{C},
\end{equation}
for some choice of norm $\|\cdot\|_C$, and we should reject the null hypothesis for large values of $T_C$.

There is no ambiguity when we have equality of the network distributions, but network distributions may be unequal in several distinct ways. For instance:
\begin{enumerate}[(I)]
\item We may have diffuse errors $X_2=X_1+E$, where the rows of $E$ independently come from some isotropic distribution on $\RR^d$.
\item We may have a rank-one error of the form $X_2=X_1 + ab^{\top}$, $a\in \RR^n, b\in \RR^d$, or a low-rank error with multiple such terms. In this case, the angle between $a$ and $\mathrm{span}(X_1)$ may be significant for determining the effect on the probability distribution.
\item We may have ``salt-and-pepper" errors where most rows of $X_2$ are the same as the corresponding rows of $X_1$, but some of the rows are changed.
\end{enumerate}
Different changes in distribution will be more likely in some contexts than others. For example, in email communication networks, a small number of employees might dramatically alter their behavior between one time window and the next, while most stay the same or only change slightly. 
On the other hand, if a new working group emerges between one time window and the next, this might result in a rank-one change to the latent positions as some team members from various groups get reassigned. The spectral norm may be more sensitive to this kind of change. See \cite{zuzul2021dynamic} for an analysis of the dynamics of such organizational communication networks.

In the present work, we consider three two-sample tests of hypothesis for the $n \times d$-dimensional latent position matrices $X_1$ and $X_2$ for a pair of random dot product networks. In all cases, the null hypothesis is that
\begin{equation*}
  H_{0}: X_1=X_2W\text{ for some }W\in \RR^{d\times d}, W^{\top} W=I.
\end{equation*}
The three alternatives we consider are
\begin{enumerate}[(I)]
\item The \emph{diffuse alternative} $$H_{A}^{\mathrm{D}}: X_2=X_1W+E,$$ where $E_i \overset{\mathrm{iid}}{\sim} \mathcal{N}(0,\sigma^2 I_d)$. In this case, the \emph{strength of the alternative} is determined by the variance of the diffuse noise in each row, $\sigma>0$.
\item The \emph{rank-one alternative with angle }$\theta$ $$H_A^{\theta}: X_2=X_1W+\sigma ab^{\top},$$ $a\in \RR^n$, $b\in \RR^d$, $\|a\|=\|b\|=1$, $\sin(\angle (\mathrm{span}(a),\mathrm{span}(X_1)))=\theta$. In this case, the \emph{strength of the alternative} is given by the size of the rank-one perturbation measured by spectral norm, $\sigma>0$.
\item The \emph{salt-and-pepper alternative} $$H_A^{\mathrm{SP}}: (X_2)_v= (X_1 W)_v\text{ for }v\in V_0, (X_2)_v \overset{\mathrm{iid}}{\sim} F\text{ for }v\in V_1.$$ Here $F$ is some distribution on $\RR^d$, and $V_0,V_1\subseteq V$ have sizes 
$$|V_0|=(1-\sigma)|V|, |V_1|=\sigma|V|.$$ The \emph{strength of the alternative} is given by the proportion of latent positions that are changed, denoted $\sigma\in[0,1]$.
\end{enumerate}

We consider the following test statistics $T_C$, where $C$ denotes the choice of norm:
\begin{enumerate}[(1)]
\item The Frobenius norm test statistic:
\begin{equation*}
    T_{\mathrm{F}} = \min_{W\in \mathbb R^{d\times d}, W^{\top}W = I} \lVert X_1^{\top} - W X_2^{\top} \rVert_{\mathrm{F}}
    = f_{\mathrm{F}}(W_{\mathrm{F}}^*)
    ;
\end{equation*}
\item The Spectral norm test statistic:
\begin{equation*}
    T_{\mathrm{S}} = \min_{W\in \mathbb R^{d\times d}, W^{\top}W = I} \lVert X_1^{\top} - W X_2^{\top} \rVert_{\mathrm{S}}
    = f_{\mathrm{S}}(W_{\mathrm{S}}^*)
    ;
\end{equation*}
\item The Robust norm test statistic:
\begin{equation*}
    T_{\mathrm{R}} = \min_{W\in \mathbb R^{d\times d}, W^{\top}W = I} \lVert X_1^{\top} - W X_2^{\top} \rVert_{\mathrm{R}}
    = f_{\mathrm{R}}(W_{\mathrm{R}}^*)
    .
\end{equation*}
\end{enumerate}
In the latter two cases, we also consider the test statistics obtained where the solution $W_{\mathrm{F}}^*$ to the Frobenius norm minimization problem is used in place of the optimal choice of orthogonal minimizer for that norm, and denote these test statistics by $\hat{T}_{\mathrm{S}} = f_{\mathrm{S}}(W_{\mathrm{F}}^*)$ and $\hat{T}_{\mathrm{R}} = f_{\mathrm{R}}(W_{\mathrm{F}}^*)$.

Because obtaining precise distributional characterizations of these test statistics is often impossible, we use a bootstrapping procedure to estimate critical values. If the true latent positions are known, we can generate repeated independent draws from random dot product graphs with these latent positions and use these to produce empirical distributions for the test statistics given above. From these empirical distributions, we can extract a bootstrapped estimate of the critical value. We describe this in Section \ref{sec:hypothesis_testing}. An analogous method can be used even when the true latent positions are unknown, as follows. Given an adjacency matrix $A_1$ for a random dot product graph, we compute its adjacency spectral embedding $\hat{X}_1$ and use this to generate a ``probability'' matrix $\hat{P}_1=\hat{X}_1\hat{X}_1^{\top}$ (with adjustments to ensure all entries are in the unit interval). Once this matrix of probabilities is generated, it is possible to simulate multiple independent network realizations based on $\hat{P}_1$. Repeating this with the second adjacency matrix $A_2$ and generating independent network realizations based on $\hat{P}_2$ makes it possible to compute multiple realizations of the appropriate test statistic, from whose empirical distribution the corresponding critical value can be estimated.  For more on bootstrapping generally, see \cite{efron1994introduction}. For specific applications to the semiparametric test, see \cite{tang2017semiparametric}. For an overview of the intricacies of bootstrapping for networks, see \cite{levin2025bootstrapping}.

\section{Two-dimensional Illustration of Spectral and Robust Procrustes}\label{sec:Duck_example}

\pgfplotstableread[col sep = comma]{data/duck/abstract-duck-orig.csv}\duck
\pgfplotstableread[col sep = comma]{data/duck/abstract-duck-rotated.csv}\duckRotated
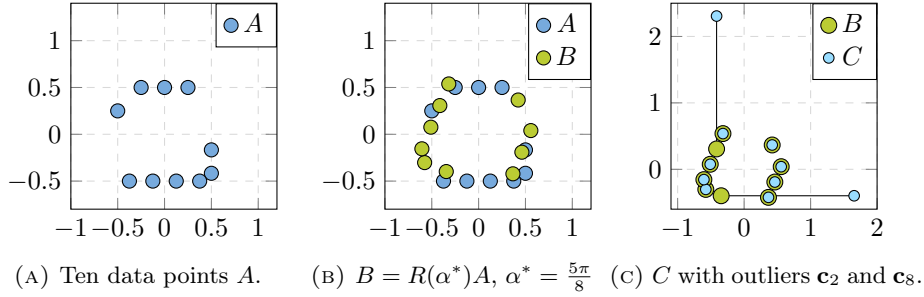
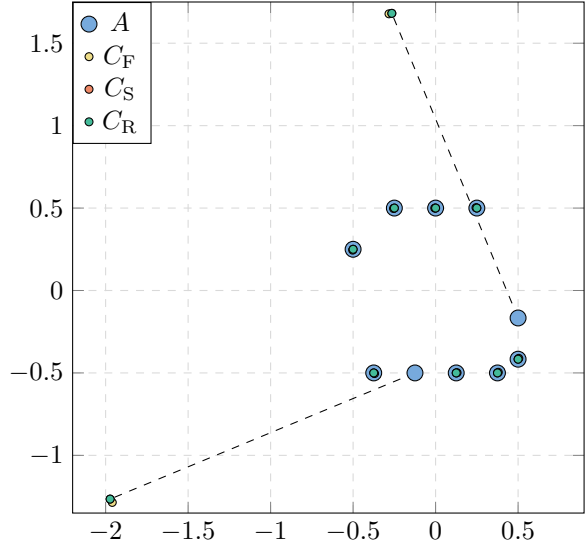
\begin{figure}
    \pgfplotstableread[col sep = comma]{data/duck/abstract-duck-even-orig-outlier1.csv}\duckOutlierOne
    \pgfplotstableread[col sep = comma]{data/duck/abstract-duck-even-orig-outlier2.csv}\duckOutlierTwo
    \pgfplotstableread[col sep = comma]{data/duck/abstract-duck-even-rotated-outliers.csv}\duckRotatedOutliers
    \pgfplotstableread[col sep = comma]{data/duck/abstract-duck-even-rotated-outlier1.csv}\duckRotatedOutlierOne
    \pgfplotstableread[col sep = comma]{data/duck/abstract-duck-even-rotated-outlier2.csv}\duckRotatedOutlierTwo
    \pgfplotstableread[col sep = comma]{data/duck/abstract-duck-even-Result-S.csv}\duckS
    \pgfplotstableread[col sep = comma]{data/duck/abstract-duck-even-Result-F.csv}\duckF
    \pgfplotstableread[col sep = comma]{data/duck/abstract-duck-even-Result-T.csv}\duckT
    \begin{subfigure}{0.33\textwidth}
        \centering
        \begin{tikzpicture}
            \begin{axis}[
            width=1.2\textwidth,
            axis equal image,
            grid = major,
            grid style={dashed, gray!30},
            legend style={at={(1,1)}, anchor=north east},
            xmin=-1.0,
            xmax=1.2,
            ymin=-0.8,
            ymax=1.4,
         ]
            \addplot [only marks,mark=*,mark options ={fill=TolLightBlue, draw=none, scale=1.33}] table [x=x, y=y, col sep=comma] {\duck};
            \addlegendentry{$A$}
            \end{axis}
        \end{tikzpicture}
        \caption{Ten data points $A$.}\label{fig:duck:orig}
    \end{subfigure}%
    \begin{subfigure}{0.33\textwidth}
        \centering
        \begin{tikzpicture}
            \begin{axis}[
            width=1.2\textwidth,
            grid = major,
            grid style={dashed, gray!30},
            legend style={at={(1,1)}, anchor=north east},
            axis equal image,
            xmin=-1.0,
            xmax=1.2,
            ymin=-0.8,
            ymax=1.4,
            ]
        \addplot [only marks,mark=*,mark options ={fill=TolLightBlue, draw=none, scale=1.33}] table [x=x, y=y, col sep=comma] {\duck};
        \addlegendentry{$A$}
        \addplot [only marks,mark=*,mark options ={fill=TolLightPear, draw=none, scale=1.33}] table [x=x, y=y, col sep=comma] {\duckRotated};
        \addlegendentry{$B$}
           \end{axis}
        \end{tikzpicture}
        \caption{$B = R(\alpha^*)A$, $\alpha^* = \frac{5\pi}{8}$}\label{fig:duck:rotate}
    \end{subfigure}%
    \begin{subfigure}{0.33\textwidth}
        \centering
        \begin{tikzpicture}
            \begin{axis}[
            width=1.2\textwidth,
            axis equal image,
            grid = major,
            grid style={dashed, gray!30},
            legend style={at={(1,1)}, anchor=north east},
            xmin=-1.1,
            xmax=2.0,
            ymin=-0.6,
            ymax= 2.5,
            ]
        \addplot [thin, black, forget plot] table [x=x, y=y, col sep=comma] {\duckRotatedOutlierOne};
        \addplot [thin, black, forget plot] table [x=x, y=y, col sep=comma] {\duckRotatedOutlierTwo};
        \addplot [only marks,mark=*,mark options ={fill=TolLightPear, draw=none, scale=1.5}] table [x=x, y=y, col sep=comma] {\duckRotated};
        \addlegendentry{$B$}
        \addplot [only marks,mark=*,mark options ={fill=TolLightCyan, draw=none, scale=1.0}] table [x=x, y=y, col sep=comma] {\duckRotatedOutliers};
        \addlegendentry{$C$}
           \end{axis}
        \end{tikzpicture}
        \caption{$C$ with outliers $\mathbf{c}_2$ and $\mathbf{c}_8$.}\label{fig:duck:outlier2}
    \end{subfigure}
    \\[2\baselineskip]
    \hspace{.026\textwidth}%
    \begin{subfigure}{0.64\textwidth}%
        \centering
        \begin{tikzpicture}
            \begin{axis}[
            width=1.2\textwidth,
            axis equal image,
            grid = major,
            grid style={dashed, gray!30},
            legend style={at={(0,1)}, anchor=north west},
            xmin=-2.2,
            xmax=0.9,
            ymin=-1.35,
            ymax=1.75,
            ]
        \addplot [thin, black, dashed, forget plot] table [x=x, y=y, col sep=comma] {\duckOutlierOne};
        \addplot [thin, black, dashed, forget plot] table [x=x, y=y, col sep=comma] {\duckOutlierTwo};
        \addplot [only marks,mark=*,mark options ={fill=TolLightBlue, draw=none, scale=1.5}] table [x=x, y=y, col sep=comma] {\duck};
        \addlegendentry{$A$}
        \addplot [only marks,mark=*,mark options ={fill=TolLightYellow, draw=none, scale=0.75}] table [x=x, y=y, col sep=comma] {\duckF};
        \addlegendentry{$C_{\mathrm{F}}$}
        \addplot [only marks,mark=*,mark options ={fill=TolLightOrange, draw=none, scale=0.75}] table [x=x, y=y, col sep=comma] {\duckS};
        \addlegendentry{$C_{\mathrm{S}}$}
        \addplot [only marks,mark=*,mark options ={fill=TolLightMint, draw=none, scale=0.75}] table [x=x, y=y, col sep=comma] {\duckT};
        \addlegendentry{$C_{\mathrm{R}}$}
        \end{axis}
        \end{tikzpicture}
        \caption{%
        The reconstructed results for Orthogonal ($C_{\mathrm{F}}$), Spectral ($C_{\mathrm{S}}$) and Robust ($C_{\mathrm{R}}$) Procrustes.
        }\label{fig:duck:result-even}
    \end{subfigure}
\caption{The two-dimensional alignment task comparing Frobenius, Spectral, and Robust Procrustes when outliers are present}
\end{figure}

For a useful illustration of the consequences of Procrustes alignments, we consider a two-dimensional example to clarify the effect of the different rotations on a set of points. 
Let the set of points be given as columns of the matrix
\begin{equation*}
A = \begin{pmatrix}
    -\frac{1}{2} & \frac{1}{2} & \frac{1}{2} & - \frac{1}{4} & 0 & \frac{1}{4} & -\frac{3}{8} & -\frac{1}{8} & \frac{1}{8} & \frac{3}{8}\\
    \frac{1}{4} & -\frac{2}{12} & -\frac{5}{12} & \frac{1}{2} & \frac{1}{2} & \frac{1}{2} & -\frac{1}{2} & -\frac{1}{2} & -\frac{1}{2} & -\frac{1}{2}
\end{pmatrix} \in \mathbb R^{2\times 10},
\end{equation*}
also shown in Figure~\ref{fig:duck:orig}. The columns of $A$ are denoted by $\mathbf{a}_1,\ldots\mathbf{a}_{10}$.
We denote by $R_\alpha = \begin{pmatrix}
    \cos\alpha&\sin\alpha\\-\sin\alpha&\cos\alpha
\end{pmatrix}$ the rotation matrix by $\alpha$ degrees and set $\alpha^*=\frac{5\pi}{8}$ as well as $W^* = R(\alpha^*)$. The rotated set of points is the matrix $B = R(\alpha^*)A$, with columns $\mathbf{b}_i$, $i=1,\ldots,10$, which is jointly visualized with the points of $A$ in Figure~\ref{fig:duck:rotate}.

For the first set of experiments, these points $\mathbf{b}_i$ are affected by outliers.
We choose two indices $\{2,8\}$ of points affected by noise and construct the set $C$ of points, whose columns are
\begin{equation*}
    \mathbf{c}_2 = \mathbf{b}_2 + \begin{pmatrix} 2\\0 \end{pmatrix}
    \qquad
    \mathbf{c}_8 = \mathbf{b}_8 + \begin{pmatrix} 0\\2 \end{pmatrix}
    \qquad
    \mathbf{c}_i = \mathbf{b}_i\quad\text{for}\quad i\notin\{2,8\}.
\end{equation*}
This is shown in Figure~\ref{fig:duck:outlier2}, where we also connect the data points $\mathbf{b}_i$ and their outliers $\mathbf{c}_i$, $i=2,8$ with a solid line.

To determine how well $C$ can be “rotated back” to $A$, let $n \in \{\mathrm{F}, \mathrm{S}, \mathrm{R}\}$
indicate the three minimization problems of (Frobenius) orthogonal Procrustes~\eqref{eq:OrthogonalProcrustes}, Spectral Procrustes~\eqref{eq:SpectralProcrustes}, and Robust Procrustes~\eqref{eq:RobustProcrustes}, respectively. For each of these, we compute the minimizers and let $C_n = W_n^*C$ be the reconstruction
of the point set and $\alpha_n$ the corresponding reconstructed angle.
For comparison, we can also exactly “reconstruct” $C^* = R(-\alpha^*)C$, illustrated in
Figure~\ref{fig:duck:result-even} by the dashed lines to the hypothetical outliers.

We introduce two error measures for the reconstructed angle: $\varepsilon_{\alpha} \coloneqq \lVert \alpha^* - \alpha_n \rVert$
and $\varepsilon_{\mathrm{F}}$, which is the Frobenius norm of $A-C_n$ omitting outlier columns. Note that both of these measures depend on the choice of $n \in \{F, S, R\}$, but we suppress this dependence in the notation for convenience. Both the Spectral $C_{\mathrm{S}}$ and the Robust $C_{\mathrm{R}}$ look indistinguishable from
$A$ or the ends of the outliers. Indeed, their result is exact up to $3\cdot10^{-11}$ in the reconstruction of the angle $\alpha^*$, as well as the Frobenius norm on the points that are not outliers.
The orthogonal Procrustes, on the other hand, has an error of $10^{-2}$ in both the angle and the Frobenius norm in the non-outliers. These results are in column $C_n$ of Table~\ref{table:outlier-errors}, for each choice of norm $n$.

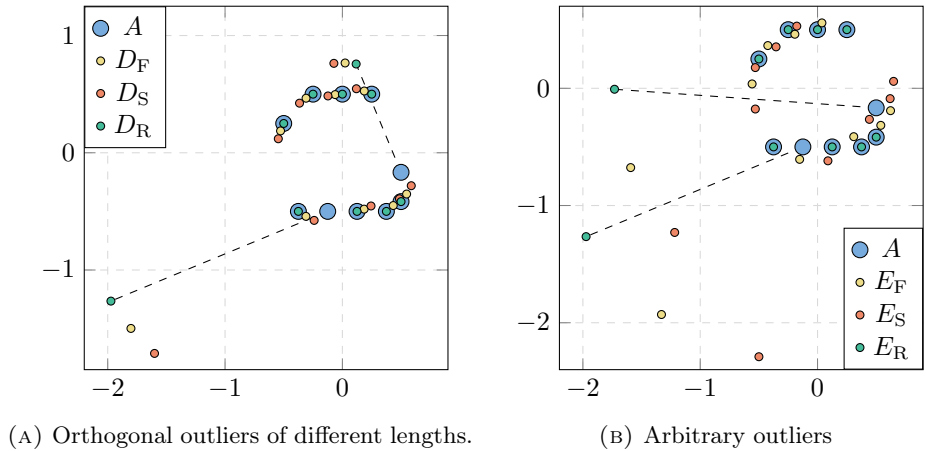
\begin{figure}
    \begin{subfigure}{0.49\textwidth}%
        \pgfplotstableread[col sep = comma]{data/duck/abstract-duck-orth-orig-outlier1.csv}\duckOrthOutlierOne
        \pgfplotstableread[col sep = comma]{data/duck/abstract-duck-orth-orig-outlier2.csv}\duckOrthOutlierTwo
        \pgfplotstableread[col sep = comma]{data/duck/abstract-duck-orth-Result-S.csv}\duckOrthS
        \pgfplotstableread[col sep = comma]{data/duck/abstract-duck-orth-Result-F.csv}\duckOrthF
        \pgfplotstableread[col sep = comma]{data/duck/abstract-duck-orth-Result-T.csv}\duckOrthT
            \centering
        \begin{tikzpicture}
            \begin{axis}[
            width=1.2\textwidth,
            axis equal image,
            grid = major,
            grid style={dashed, gray!30},
            legend style={at={(0,1)}, anchor=north west},
            xmin=-2.2,
            xmax=0.9,
            ymin=-1.85,
            ymax=1.25,
            ]
        \addplot [thin, black, dashed, forget plot] table [x=x, y=y, col sep=comma] {\duckOrthOutlierOne};
        \addplot [thin, black, dashed, forget plot] table [x=x, y=y, col sep=comma] {\duckOrthOutlierTwo};
        \addplot [only marks,mark=*,mark options ={fill=TolLightBlue, draw=none, scale=1.5}] table [x=x, y=y, col sep=comma] {\duck};
        \addlegendentry{$A$}
        \addplot [only marks,mark=*,mark options ={fill=TolLightYellow, draw=none, scale=0.75}] table [x=x, y=y, col sep=comma] {\duckOrthF};
        \addlegendentry{$D_{\mathrm{F}}$}
        \addplot [only marks,mark=*,mark options ={fill=TolLightOrange, draw=none, scale=0.75}] table [x=x, y=y, col sep=comma] {\duckOrthS};
        \addlegendentry{$D_{\mathrm{S}}$}
        \addplot [only marks,mark=*,mark options ={fill=TolLightMint, draw=none, scale=0.75}] table [x=x, y=y, col sep=comma] {\duckOrthT};
        \addlegendentry{$D_{\mathrm{R}}$}
        \end{axis}
        \end{tikzpicture}
        \caption{%
        Orthogonal outliers of different lengths.
        }\label{fig:duck:result-orth}
    \end{subfigure}
    \begin{subfigure}{0.49\textwidth}%
        \pgfplotstableread[col sep = comma]{data/duck/abstract-duck-any-orig-outlier1.csv}\duckAnyOutlierOne
        \pgfplotstableread[col sep = comma]{data/duck/abstract-duck-any-orig-outlier2.csv}\duckAnyOutlierTwo
        \pgfplotstableread[col sep = comma]{data/duck/abstract-duck-any-Result-S.csv}\duckAnyS
        \pgfplotstableread[col sep = comma]{data/duck/abstract-duck-any-Result-F.csv}\duckAnyF
        \pgfplotstableread[col sep = comma]{data/duck/abstract-duck-any-Result-T.csv}\duckAnyT
            \centering
        \begin{tikzpicture}
            \begin{axis}[
            width=1.2\textwidth,
            axis equal image,
            grid = major,
            grid style={dashed, gray!30},
            legend style={at={(1,0)}, anchor=south east},
            xmin=-2.2,
            xmax=0.9,
            ymin=-2.4,
            ymax=0.7,
            ]
        \addplot [thin, black, dashed, forget plot] table [x=x, y=y, col sep=comma] {\duckAnyOutlierOne};
        \addplot [thin, black, dashed, forget plot] table [x=x, y=y, col sep=comma] {\duckAnyOutlierTwo};
        \addplot [only marks,mark=*,mark options ={fill=TolLightBlue, draw=none, scale=1.5}] table [x=x, y=y, col sep=comma] {\duck};
        \addlegendentry{$A$}
        \addplot [only marks,mark=*,mark options ={fill=TolLightYellow, draw=none, scale=0.75}] table [x=x, y=y, col sep=comma] {\duckAnyF};
        \addlegendentry{$E_{\mathrm{F}}$}
        \addplot [only marks,mark=*,mark options ={fill=TolLightOrange, draw=none, scale=0.75}] table [x=x, y=y, col sep=comma] {\duckAnyS};
        \addlegendentry{$E_{\mathrm{S}}$}
        \addplot [only marks,mark=*,mark options ={fill=TolLightMint, draw=none, scale=0.75}] table [x=x, y=y, col sep=comma] {\duckAnyT};
        \addlegendentry{$E_{\mathrm{R}}$}
        \end{axis}
        \end{tikzpicture}
        \caption{%
        Arbitrary outliers
        }\label{fig:duck:result-any}
    \end{subfigure}
\caption{Orthogonal and arbitrary outliers.}\label{fig:duck:result-orth-and-any}
\end{figure}

For our second and third examples, let $D$ be a matrix of points with columns $\mathbf{d}_i$, $i=1,\ldots,10$, where 
we keep the outliers orthogonal but of different lengths
\begin{equation*}
    \mathbf{d}_2 = \mathbf{b}_2 + \begin{pmatrix} 1\\0 \end{pmatrix}
    \qquad
    \mathbf{d}_8 = \mathbf{b}_8 + \begin{pmatrix} 0\\2 \end{pmatrix}
    \qquad
    \mathbf{d}_i = \mathbf{b}_i\quad\text{for}\quad i\notin\{2,8\}.
\end{equation*}
For the third set of points, we construct a matrix $E$ with outliers that are neither of same length nor orthogonal
\begin{equation*}
    \mathbf{e}_2 = \mathbf{b}_2 + \begin{pmatrix} 1\\2 \end{pmatrix}
    \qquad
    \mathbf{e}_8 = \mathbf{b}_8 + \begin{pmatrix} 0\\2 \end{pmatrix}
    \qquad
    \mathbf{e}_i = \mathbf{b}_i\quad\text{for}\quad i\notin\{2,8\}.
\end{equation*}

Both situations are depicted in Figure~\ref{fig:duck:result-orth-and-any},
their errors in Table~\ref{table:outlier-errors}.
We see that Orthogonal Procrustes and Spectral Procrustes reconstruct increasingly poorly, with Spectral worsening faster when moving away from the orthonormal outlier scenario given by matrix $C$. Robust Procrustes, by contrast, still reconstructs these outlier scenarios up to machine precision.

\begin{table}[tbp]
    \caption{Errors in angle ($\varepsilon_\alpha$) and Frobenius on the non-outlier points ($\varepsilon_{\mathrm{F}}$) for the sets $C$, $D$, and $E$ for all three estimators 
    Orthogonal (F), Spectral (S) and Robust (R) Procrustes.}
    \label{table:outlier-errors}
    \sisetup{
        table-alignment-mode = format,
        table-number-alignment = right,
        table-format = 1.1e2,
        round-mode = places,
        round-precision = 2,
        exponent-product = \cdot
        }
    \begin{tabular}{rSSSSSS}
        \toprule
        {$n$} & \multicolumn{2}{c}{$C_n$} & \multicolumn{2}{c}{$D_n$} & \multicolumn{2}{c}{$E_n$}
        \\\cmidrule(r){2-3}\cmidrule(lr){4-5}\cmidrule(l){6-7}
        & $\varepsilon_{\alpha}$ & $\varepsilon_{\mathrm{F}}$ & $\varepsilon_{\alpha}$ & $\varepsilon_{\mathrm{F}}$ & $\varepsilon_{\alpha}$ & $\varepsilon_{\mathrm{F}}$\\
        \midrule
        {$\mathrm{F}$} & 0.010 & 0.0166399 & 0.123315 & 0.200917 & 0.785612 & 1.24812
        \\
        {$\mathrm{S}$} & 3.435e-11 & 5.59975e-11 & 0.248392 & 0.403922 & 0.3969491 & 0.642918 \\
        {$\mathrm{R}$} & 4.294e-11 & 7.00052e-11 & 2.80183e-11 & 4.56788e-11 & 3.39313e-11 & 5.53193e-11 \\
        \bottomrule
    \end{tabular}
\end{table}

\begin{figure}
    \begin{subfigure}{0.49\textwidth}%
        \pgfplotstableread[col sep = comma]{data/duck/abstract-duck-spectral-orig-outlier1.csv}\duckSpecOutlierOne
        \pgfplotstableread[col sep = comma]{data/duck/abstract-duck-spectral-orig-outlier2.csv}\duckSpecOutlierTwo
        \pgfplotstableread[col sep = comma]{data/duck/abstract-duck-spectral-orig-outlier3.csv}\duckSpecOutlierThree
        \pgfplotstableread[col sep = comma]{data/duck/abstract-duck-spectral-orig-outlier4.csv}\duckSpecOutlierFour
        \pgfplotstableread[col sep = comma]{data/duck/abstract-duck-spectral-orig-outlier5.csv}\duckSpecOutlierFive
        \pgfplotstableread[col sep = comma]{data/duck/abstract-duck-spectral-orig-outlier6.csv}\duckSpecOutlierSix
        \pgfplotstableread[col sep = comma]{data/duck/abstract-duck-spectral-orig-outlier7.csv}\duckSpecOutlierSeven
        \pgfplotstableread[col sep = comma]{data/duck/abstract-duck-spectral-orig-outlier8.csv}\duckSpecOutlierEight
        \pgfplotstableread[col sep = comma]{data/duck/abstract-duck-spectral-orig-outlier9.csv}\duckSpecOutlierNine
        \pgfplotstableread[col sep = comma]{data/duck/abstract-duck-spectral-orig-outlier10.csv}\duckSpecOutlierTen
        \pgfplotstableread[col sep = comma]{data/duck/abstract-duck-spectral-Result-S.csv}\duckSpecS
        \pgfplotstableread[col sep = comma]{data/duck/abstract-duck-spectral-Result-F.csv}\duckSpecF
        \pgfplotstableread[col sep = comma]{data/duck/abstract-duck-spectral-Result-T.csv}\duckSpecT
            \centering
        \begin{tikzpicture}
            \begin{axis}[
            width=1.2\textwidth,
            axis equal image,
            grid = major,
            grid style={dashed, gray!30},
            legend style={at={(0,0)}, anchor=south west},
            xmin=-2.4,
            xmax=0.7,
            ymin=-1.7,
            ymax=1.4,
            ]
        \addplot [thin, black, dashed, forget plot] table [x=x, y=y, col sep=comma] {\duckSpecOutlierOne};
        \addplot [thin, black, dashed, forget plot] table [x=x, y=y, col sep=comma] {\duckSpecOutlierTwo};
        \addplot [thin, black, dashed, forget plot] table [x=x, y=y, col sep=comma] {\duckSpecOutlierThree};
        \addplot [thin, black, dashed, forget plot] table [x=x, y=y, col sep=comma] {\duckSpecOutlierFour};
        \addplot [thin, black, dashed, forget plot] table [x=x, y=y, col sep=comma] {\duckSpecOutlierFive};
        \addplot [thin, black, dashed, forget plot] table [x=x, y=y, col sep=comma] {\duckSpecOutlierSix};
        \addplot [thin, black, dashed, forget plot] table [x=x, y=y, col sep=comma] {\duckSpecOutlierSeven};
        \addplot [thin, black, dashed, forget plot] table [x=x, y=y, col sep=comma] {\duckSpecOutlierEight};@
        \addplot [thin, black, dashed, forget plot] table [x=x, y=y, col sep=comma] {\duckSpecOutlierNine};
        \addplot [thin, black, dashed, forget plot] table [x=x, y=y, col sep=comma] {\duckSpecOutlierTen};
        \addplot [only marks,mark=*,mark options ={fill=TolLightBlue, draw=none, scale=1.5}] table [x=x, y=y, col sep=comma] {\duck};
        \addlegendentry{$A$}
        \addplot [only marks,mark=*,mark options ={fill=TolLightYellow, draw=none, scale=0.75}] table [x=x, y=y, col sep=comma] {\duckSpecF};
        \addlegendentry{$\tilde C_{\mathrm{F}}$}
        \addplot [only marks,mark=*,mark options ={fill=TolLightOrange, draw=none, scale=0.75}] table [x=x, y=y, col sep=comma] {\duckSpecS};
        \addlegendentry{$\tilde C_{\mathrm{S}}$}
        \addplot [only marks,mark=*,mark options ={fill=TolLightMint, draw=none, scale=0.75}] table [x=x, y=y, col sep=comma] {\duckSpecT};
        \addlegendentry{$\tilde C_{\mathrm{R}}$}
        \end{axis}
        \end{tikzpicture}
        \caption{%
        Addition of an element from $\mathcal N(B)$.
        \\\ 
        }\label{fig:duck:result-spec}
    \end{subfigure}
    \begin{subfigure}{0.49\textwidth}%
        \pgfplotstableread[col sep = comma]{data/duck/abstract-duck-spectral_and_uniform-orig-outlier1.csv}\duckSpecGaussOutlierOne
        \pgfplotstableread[col sep = comma]{data/duck/abstract-duck-spectral_and_uniform-orig-outlier2.csv}\duckSpecGaussOutlierTwo
        \pgfplotstableread[col sep = comma]{data/duck/abstract-duck-spectral_and_uniform-orig-outlier3.csv}\duckSpecGaussOutlierThree
        \pgfplotstableread[col sep = comma]{data/duck/abstract-duck-spectral_and_uniform-orig-outlier4.csv}\duckSpecGaussOutlierFour
        \pgfplotstableread[col sep = comma]{data/duck/abstract-duck-spectral_and_uniform-orig-outlier5.csv}\duckSpecGaussOutlierFive
        \pgfplotstableread[col sep = comma]{data/duck/abstract-duck-spectral_and_uniform-orig-outlier6.csv}\duckSpecGaussOutlierSix
        \pgfplotstableread[col sep = comma]{data/duck/abstract-duck-spectral_and_uniform-orig-outlier7.csv}\duckSpecGaussOutlierSeven
        \pgfplotstableread[col sep = comma]{data/duck/abstract-duck-spectral_and_uniform-orig-outlier8.csv}\duckSpecGaussOutlierEight
        \pgfplotstableread[col sep = comma]{data/duck/abstract-duck-spectral_and_uniform-orig-outlier9.csv}\duckSpecGaussOutlierNine
        \pgfplotstableread[col sep = comma]{data/duck/abstract-duck-spectral_and_uniform-orig-outlier10.csv}\duckSpecGaussOutlierTen
        \pgfplotstableread[col sep = comma]{data/duck/abstract-duck-spectral_and_uniform-Result-S.csv}\duckSpecGaussS
        \pgfplotstableread[col sep = comma]{data/duck/abstract-duck-spectral_and_uniform-Result-F.csv}\duckSpecGaussF
        \pgfplotstableread[col sep = comma]{data/duck/abstract-duck-spectral_and_uniform-Result-T.csv}\duckSpecGaussT
            \centering
        \begin{tikzpicture}
            \begin{axis}[
            width=1.2\textwidth,
            axis equal image,
            grid = major,
            grid style={dashed, gray!30},
            legend style={at={(0,0)}, anchor=south west},
            xmin=-2.4,
            xmax=0.7,
            ymin=-1.7,
            ymax=1.4,
            ]
        \addplot [thin, black, dashed, forget plot] table [x=x, y=y, col sep=comma] {\duckSpecGaussOutlierOne};
        \addplot [thin, black, dashed, forget plot] table [x=x, y=y, col sep=comma] {\duckSpecGaussOutlierTwo};
        \addplot [thin, black, dashed, forget plot] table [x=x, y=y, col sep=comma] {\duckSpecGaussOutlierThree};
        \addplot [thin, black, dashed, forget plot] table [x=x, y=y, col sep=comma] {\duckSpecGaussOutlierFour};
        \addplot [thin, black, dashed, forget plot] table [x=x, y=y, col sep=comma] {\duckSpecGaussOutlierFive};
        \addplot [thin, black, dashed, forget plot] table [x=x, y=y, col sep=comma] {\duckSpecGaussOutlierSix};
        \addplot [thin, black, dashed, forget plot] table [x=x, y=y, col sep=comma] {\duckSpecGaussOutlierSeven};
        \addplot [thin, black, dashed, forget plot] table [x=x, y=y, col sep=comma] {\duckSpecGaussOutlierEight};
        \addplot [thin, black, dashed, forget plot] table [x=x, y=y, col sep=comma] {\duckSpecGaussOutlierNine};
        \addplot [thin, black, dashed, forget plot] table [x=x, y=y, col sep=comma] {\duckSpecGaussOutlierTen};
        \addplot [only marks,mark=*,mark options ={fill=TolLightBlue, draw=none, scale=1.5}] table [x=x, y=y, col sep=comma] {\duck};
        \addlegendentry{$A$}
        \addplot [only marks,mark=*,mark options ={fill=TolLightYellow, draw=none, scale=0.75}] table [x=x, y=y, col sep=comma] {\duckSpecGaussF};
        \addlegendentry{$\tilde D_{\mathrm{F}}$}
        \addplot [only marks,mark=*,mark options ={fill=TolLightOrange, draw=none, scale=0.75}] table [x=x, y=y, col sep=comma] {\duckSpecGaussS};
        \addlegendentry{$\tilde D_{\mathrm{S}}$}
        \addplot [only marks,mark=*,mark options ={fill=TolLightMint, draw=none, scale=0.75}] table [x=x, y=y, col sep=comma] {\duckSpecGaussT};
        \addlegendentry{$\tilde D_{\mathrm{R}}$}
        \end{axis}
        \end{tikzpicture}
        \caption{Addition of an element from $\mathcal N(B)$, plus additional uniform noise.
        }\label{fig:duck:result-spec-uniform}
    \end{subfigure}
\caption{Two global dislocations affecting all points.}\label{fig:duck:result-twospec}
\end{figure}
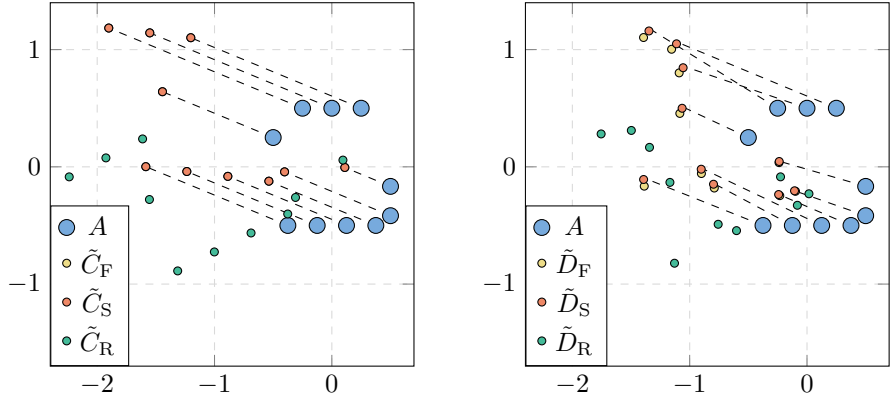

For the outlier scenarios corresponding to the matrices $C$, $D$, and $E$ of the first set of experiments, outliers are well reconstructed by the robust choice. Thus in the second set of experiments, we consider a global error on $B$, represented with dashed lines, and again illustrate the results of the reconstruction with different norms.

First, we construct data sets $\tilde C$ and $\tilde D$ by considering the null space $\mathcal N(B) = \{ \mathbf{v} \in \mathbb R^{10}, B\mathbf{v} = \mathbf{0}\}$, which is eight-dimensional. Let $\mathbf{v}_1,\ldots\mathbf{v}_8$ be an orthonormal basis of $\mathcal N(b)$. Define $\mathbf{v} = \sum_{i=1}^8 \mathbf{v}_i$ and $\mathbf{u} = \begin{pmatrix}
    1\\1
\end{pmatrix}$ and then generate 
\begin{equation*}
    \tilde C = B + \mathbf{u}\mathbf{v}^{\top}
    \qquad\text{ and }\qquad \tilde D = B + \frac{1}{2}\mathbf{u}\mathbf{v}^{\top} + R,
\end{equation*}
where $R$ is a matrix with random \iid uniform noise on $[0,\frac{1}{2}]$.

Figure \ref{fig:duck:result-twospec} illustrates the corresponding scenario. The dataset $A$ with the dislocation introduced in $\tilde C, \tilde D$ is already rotated back and depicted by the dashed lines. Note that here, all points are affected by a global, structured dislocation.
We then again solve all three Procrustes problems, with the orthogonal (or Frobenius) Procrustes reconstruction $\tilde C_{\mathrm{F}}, \tilde D_{\mathrm{F}}$, spectral $\tilde C_{\mathrm{S}}, \tilde D_{\mathrm{S}}$, and robust $\tilde C_{\mathrm{R}}, \tilde D_{\mathrm{R}}$. Since all points are moved, here we can only consider the reconstructed angle error $\varepsilon_{\alpha} \coloneqq \lVert \alpha^* - \alpha_n \rVert$ as a measure. The results are given in Table \ref{table:spectral-errors}.
\begin{table}[tbp]
    \caption{Errors in angle ($\varepsilon_\alpha$) for the sets $\tilde C$, $\tilde D$ for all three Orthogonal (F), Spectral (S) and Robust (R) Procrustes}
    \label{table:spectral-errors}
    \sisetup{
        table-alignment-mode = format,
        table-number-alignment = right,
        table-format = 1.3e2,
        round-mode = places,
        round-precision = 4,
        exponent-product = \cdot
        }
    \begin{tabular}{rSS}
        \toprule
        {$n$} & {$\tilde C_n$} &  {$\tilde D_n$} \\
        \midrule
        {$\mathrm{F}$} & 2.22045e-16 & 0.0516324\\
        {$\mathrm{S}$} & 1.6904e-8 & 0.0107992\\
        {$\mathrm{R}$} & 0.595122 & 0.562687\\
        \bottomrule
    \end{tabular}
\end{table}
For $\tilde C$, the reconstructions of orthogonal $\tilde C_{\mathrm{F}}$ and spectral $\tilde C_{\mathrm{S}}$ results are visually indistinguishable in Figure~\ref{fig:duck:result-spec}. The result puts all points at the position $A+R(-\alpha^*)\mathbf{u}\mathbf{v}^\top$, which would be the original locations of the points in $A$ if the (rotated) element of the nullspace $\mathcal{N}(B)$ were added. The same can be seen in $\varepsilon_\alpha \coloneqq \lVert \alpha^* - \alpha_n \rVert$ in Table~\ref{table:spectral-errors}, where the closed-form solution of the Orthogonal Procrustes is marginally closer than the spectral one obtained by LTMADS, which is an iterative solver.
Compared to that, the Robust Procrustes does not reconstruct well, neither visually nor in the angle.

For $\tilde D$, we see that Orthogonal Procrustes is now off the mark, while Spectral Procrustes still gets reasonably close to the true angle. This can also be seen in Figure~\ref{fig:duck:result-spec-uniform}, where the yellow dots of $\tilde D_{\mathrm{F}}$ are slightly off of their expected positions at the ends of the dashed lines. The Spectral result $\tilde D_{\mathrm{S}}$ appears right on target. Again, due to the global error, the Robust Procrustes cannot reconstruct the rotation angle for this case either.

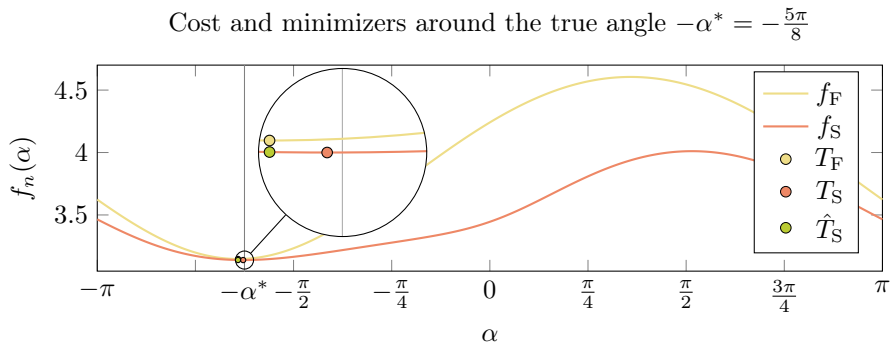
\begin{figure}
    \centering
    \pgfplotstableread[col sep = comma]{data/duck/abstract-duck-spectral_and_uniform-cost-minimisers.csv}\duckSUmin
    \pgfplotstableread[col sep = comma]{data/duck/abstract-duck-spectral_and_uniform-cost-data.csv}\duckSUcost
        \begin{tikzpicture}
    \begin{axis}[
        name=main,
        xlabel={$\alpha$},
        ylabel={$f_n(\alpha)$},
        width=.95\textwidth, height=.25\textheight,
        axis equal image,
        title={Cost and minimizers around the true angle $-\alpha^* = -\frac{5\pi}{8}$},
        ymin=3.05, ymax=4.7,
        legend pos=north east,
        legend style={mark options={mark size=15pt}},
        xmin=-pi, xmax=pi,
        xtick={-pi, -3*pi/4, -5*pi/8, -pi/2, -pi/4, 0, pi/4, pi/2, 3*pi/4, pi},
        xticklabels={$-\pi$,
        $\,$,
        $-\alpha^*$, $-\frac{\pi}{2}$, $-\frac{\pi}{4}$, $0$, $\frac{\pi}{4}$, $\frac{\pi}{2}$, $\frac{3\pi}{4}$, $\pi$},
        enlargelimits=false,
        clip=true,
    ]
        \addplot[draw, TolLightYellow, thick] table[x=x, y=F, col sep=comma]{\duckSUcost};
        \addlegendentry{$f_{\mathrm{F}}$}
        \addplot[draw, TolLightOrange, thick] table[x=x, y=S, col sep=comma]{\duckSUcost};
        \addlegendentry{$f_{\mathrm{S}}$}
        \addplot[black!50, forget plot] coordinates {(-1.9634954084936207,2) (-1.9634954084936207,5)};
        \addplot[only marks,mark=*, forget plot, mark options={fill=TolLightYellow, draw=none, scale=0.5}] table [x=xF, y=yF, col sep=comma] {\duckSUmin};
        \addlegendimage{draw=white, mark=*, mark options={mark size=2pt,fill=TolLightYellow, draw=black, line width=0.25pt}};
        \addlegendentry{$T_{\mathrm{F}}$}
        \addplot[only marks,forget plot, mark=*,mark options={fill=TolLightOrange, draw=none, scale=0.5}] table [x=xS, y=yS, col sep=comma] {\duckSUmin};
        \addlegendimage{draw=white, mark=*, mark options={mark size=2pt,fill=TolLightOrange, draw=black, line width=0.25pt}};
        \addlegendentry{$T_{\mathrm{S}}$}
        \addplot[only marks,forget plot, mark=*,mark options={fill=TolLightPear, draw=none, scale=0.5}] table [x=xF, y=yFinS, col sep=comma] {\duckSUmin};
        \addlegendimage{draw=white, mark=*, mark options={mark size=2pt,fill=TolLightPear, draw=black, line width=0.25pt}};
        \addlegendentry{$\hat T_{\mathrm{S}}$}

        \draw[black] (axis cs:-5*pi/8, 3.138810590988994) circle [radius=0.12cm];
        \draw[black, shorten <=0.12cm] (axis cs:-5*pi/8, 3.138810590988994) -- (axis cs:-3*pi/8, 4);
        \node[coordinate] (inset-center) at (axis cs:-3*pi/8, 4) {};
    \end{axis}

    \newcommand{\zoomradius}{0.06}        
    \newcommand{\zoomrange}{0.06}         
    \newcommand{\insetradius}{1.125}       

    \begin{scope}
        \fill[white] (inset-center) circle [radius=\insetradius cm];
        \clip (inset-center) circle [radius=\insetradius cm];

        \begin{axis}[
            at={(inset-center)}, anchor=center,
            width={2*\insetradius cm}, height={2*\insetradius cm},
            xmin={-5*pi/8 - \zoomrange}, xmax={-5*pi/8 + \zoomrange},
            ymin={3.138810590988994 - \zoomrange}, ymax={3.138810590988994 + \zoomrange},
            axis lines=none,
            enlargelimits=false,
            scale only axis,
        ]
            \addplot[black!33, forget plot] coordinates {(-1.9634954084936207,{3.138810590988994-\zoomrange}) (-1.9634954084936207,{3.138810590988994+\zoomrange})};
            \addplot[draw, TolLightYellow, thick] table[x=x, y=F, col sep=comma]{\duckSUcost};
            \addplot[draw, TolLightOrange, thick] table[x=x, y=S, col sep=comma]{\duckSUcost};
            \addplot[only marks,mark=*,mark options={fill=TolLightYellow, draw=none, scale=1}] table [x=xF, y=yF, col sep=comma] {\duckSUmin};
            \addplot[only marks,mark=*,mark options={fill=TolLightOrange, draw=none, scale=1}] table [x=xS, y=yS, col sep=comma] {\duckSUmin};
            \addplot[only marks,mark=*,mark options={fill=TolLightPear, draw=none, scale=1}] table [x=xF, y=yFinS, col sep=comma] {\duckSUmin};
        \end{axis}
        \draw[black, thick] (inset-center) circle [radius=\insetradius cm];
    \end{scope}
\end{tikzpicture}
    \caption{The error in angle for the experiment $\tilde D$ visualized on the corresponding cost functions $f_n$ for the case of the Frobenius ($n=\mathrm{F}$) and Spectral ($n=\mathrm{S}$) as well as the cost of the Frobenius minimizer plugged into the spectral cost.}
    \label{fig:costTildeD}
\end{figure}

We consider a third visualization of the difference between choices of norm in Figure~\ref{fig:costTildeD}, where the Frobenius cost $f_{\mathrm{F}}$ and the Spectral cost $f_{\mathrm{S}}$ are plotted around the perfect reconstruction at $-\alpha^*$. Again, the minimizer $W_{\mathrm{S}}^*$ is closer to the true angle. This figure reinforces that the cost $\hat T_{\mathrm{S}} = f_{\mathrm{S}}(W_{\mathrm{F}}^*)$ of the Frobenius minimizer in the spectral norm is even closer to the actual spectral minimum $T_{\mathrm{S}} = f_{\mathrm{S}}(W_{\mathrm{S}}^*)$.
Their difference is actually just 
$3.3342\times10^{-4}$, since both costs are similarly close to the spectral minimizer.

\section{Hypothesis Testing and Power}\label{sec:hypothesis_testing}

In this section, we consider tests of hypothesis for the semiparametric problem of testing equality of graph distributions for RDPGs under the three alternative hypotheses described in Section~\ref{section:statistical}, and using the five test statistics $T_F, T_S, T_R, \hat{T}_S,$ and $\hat{T}_R$, where the first three statistics come from Eq.~\eqref{eq:teststat} with Frobenius, Spectral, and Robust norms chosen, and $\hat{T}_S, \hat{T}_R$ come from using the corresponding choice of norm to measure the distance between matrices, but using the best Frobenius-norm orthogonal matrix $p$ in place of the optimal one for that choice of norm.

 Throughout this section, we generate latent position matrices and adjacency matrices, compute adjacency spectral embeddings find the Frobenius-norm Procrustes solutions, and implement parallelization for constructing the bootstrapped power curves using the open-source statistical software \texttt{R}. To compute the solutions to the Spectral Procrustes and Robust Procrustes problems, we again use \texttt{Julia} as described in the previous section. The translation between languages is facilitated by the use of \texttt{JuliaCall} \cite{JuliaCall}, which seamlessly integrates these optimization procedures with the \texttt{R} computing environment. Compared to only using the Frobenius-norm minimizer, we find that utilizing \texttt{JuliaCall} to find the spectral- and robust-norm minimizers increased the length of time to generate a single power curve by approximately 15\%, a modest addition that will typically become relevant only when doing thousands of alignments. For example, generating each power curve in the experiments below requires generating 42,000 adjacency matrices of size 1000$\times$1000, computing their largest eigenvectors, and aligning these matrices. Using only the Frobenius minimizer consistently takes about 1045 seconds to generate one power curve. When using the optimal minimizers for spectral and robust norms computed with \texttt{JuliaCall}, each power curve consistently takes about 1200 seconds (when both are run on a 14-core M4 MacBook Pro with 48 GB RAM). All code for these experiments can be found at \url{https://github.com/zlubberts/SpectralProcrustes}.

\subsection{Diffuse alternative}


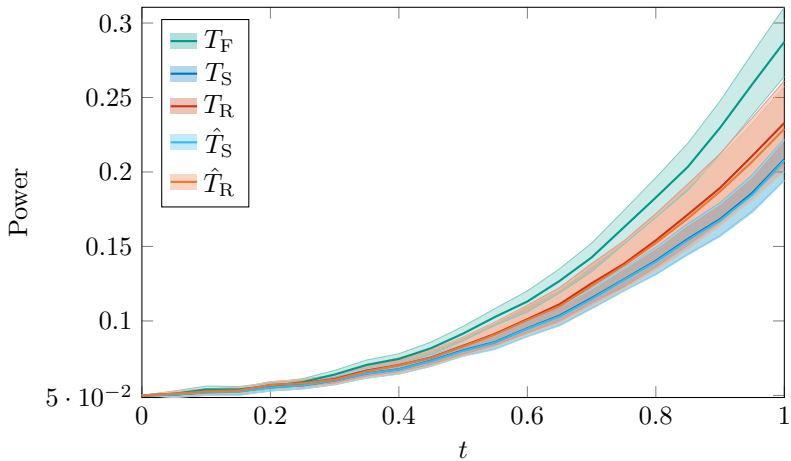
\begin{figure}
    \centering
    \begin{tikzpicture}
          \begin{axis}[
    QuantilesLegend,
    xlabel={$t$}, ylabel={Power},
    width=.8\textwidth, height=.33\textheight,
    legend pos=north west,
    enlarge x limits=false,
    enlarge y limits=false
  ]
  \addplot[name path=upper1, draw, very thin,forget plot, TF!50] table[x=t, y=u, col sep=comma]{data/powercurves_csvs/fig1curve_F.csv};
  \addplot[name path=lower1, draw, very thin,forget plot, TF!50] table[x=t, y=l, col sep=comma]{data/powercurves_csvs/fig1curve_F.csv};
  \addplot[TF, forget plot, fill opacity=0.2] fill between[of=upper1 and lower1];
  \addplot[name path=upper2, draw, very thin,forget plot, TS!50] table[x=t, y=u, col sep=comma]{data/powercurves_csvs/fig1curve_SS.csv};
  \addplot[name path=lower2, draw, very thin,forget plot, TS!50] table[x=t, y=l, col sep=comma]{data/powercurves_csvs/fig1curve_SS.csv};
  \addplot[TS, forget plot, fill opacity=0.2] fill between[of=upper2 and lower2];
  \addplot[name path=upper3, draw, very thin,forget plot, TR!50] table[x=t, y=u, col sep=comma]{data/powercurves_csvs/fig1curve_RR.csv};
  \addplot[name path=lower3, draw, very thin,forget plot, TR!50] table[x=t, y=l, col sep=comma]{data/powercurves_csvs/fig1curve_RR.csv};
  \addplot[TR, forget plot, fill opacity=0.2] fill between[of=upper3 and lower3];
  \addplot[name path=upper4, draw, very thin,forget plot, TSF!50] table[x=t, y=u, col sep=comma]{data/powercurves_csvs/fig1curve_SF.csv};
  \addplot[name path=lower4, draw, very thin,forget plot, TSF!50] table[x=t, y=l, col sep=comma]{data/powercurves_csvs/fig1curve_SF.csv};
  \addplot[TSF, forget plot, fill opacity=0.2] fill between[of=upper4 and lower4];
  \addplot[name path=upper5, draw, very thin,forget plot, TRF!50] table[x=t, y=u, col sep=comma]{data/powercurves_csvs/fig1curve_RF.csv};
  \addplot[name path=lower5, draw, very thin,forget plot, TRF!50] table[x=t, y=l, col sep=comma]{data/powercurves_csvs/fig1curve_RF.csv};
  \addplot[TRF, forget plot, fill opacity=0.2] fill between[of=upper5 and lower5];
  \addplot[thick, TF] table[x=t, y=c, col sep=comma]{data/powercurves_csvs/fig1curve_F.csv};
  \addlegendentry{$T_{\mathrm{F}}$}
  \addplot[thick, TS] table[x=t, y=c, col sep=comma]{data/powercurves_csvs/fig1curve_SS.csv};
  \addlegendentry{$T_{\mathrm{S}}$}
  \addplot[thick, TR] table[x=t, y=c, col sep=comma]{data/powercurves_csvs/fig1curve_RR.csv};
  \addlegendentry{$T_{\mathrm{R}}$}
  \addplot[thick, TSF] table[x=t, y=c, col sep=comma]{data/powercurves_csvs/fig1curve_SF.csv};
  \addlegendentry{$\hat T_{\mathrm{S}}$}
  \addplot[thick, TRF] table[x=t, y=c, col sep=comma]{data/powercurves_csvs/fig1curve_RF.csv};
  \addlegendentry{$\hat T_{\mathrm{R}}$}
  \end{axis}
    \end{tikzpicture}
    \caption{Power curves for diffuse noise in dimension 3}
    \label{fig:diffuse-noise}
\end{figure}



This setting considers the diffuse alternative (I) with the matrix $X$ corresponding to the latent positions of a {\em stochastic blockmodel} random graph with $n=1000$ vertices and $r=3$ clusters. As with random dot product graphs, stochastic blockmodel (SBM) \cite{hoff_raftery_handcock} random graphs also have connection probabilities determined by low-dimensional features associated to each vertex. For an (undirected) SBM, each vertex is assigned a community label $\tau(i)\in\{1,\ldots, r\}$; the edges arise independently; and the probability of an edge $\{u,v\}$ forming is given by $B_{\tau(u),\tau(v)}$, for a symmetric matrix of block connection probabilities $B\in \mathbb{R}^{r\times r}$. The SBM graphs we will generate all have low-rank, positive definite matrices of connection probabilities, and for this reason are also RDPGs. 

For each of the $n_{\mathrm{MC}}=100$ power curves generated, we create a stochastic blockmodel by choosing the cluster memberships from a multinomial distribution with equal probabilities for the three clusters. This means that the exact number of vertices in each block may differ from one power curve to another, but the block sizes should be approximately equal for each curve. These community memberships are recorded in the matrix $Z\in\{0,1\}^{n\times r}$, with $Z_{ia}=1$ implying that vertex $i$ is in cluster $a$, and exactly on entry of each row of $Z$ equals 1. The block connection probability matrix $B$ is drawn for each curve by drawing independent, identically distributed Uniform($0,1/r$) random variables for all of the strictly upper triangular entries, then setting the diagonal entries as \iid Uniform($0,1-1/r$). We set $X=U|S|^{1/2}$ for the eigendecomposition of the SBM probability matrix $P=\frac{1}{\sqrt{n}}Z|B|Z^{\top}=USU^{\top}$, where $|B|$ denotes the matrix $B$ with all eigenvalues having absolute value applied. For a given power curve, the matrix $Y$ is generated as $X+E$, where $E_i$ are \iid $\mathcal{N}\left(0,\frac{1}{2\sqrt{n}}I\right)$; to get the intermediate matrices on the power curve, we consider $Y_t=(1-t)X+tY$, so that $t=0$ corresponds to the null hypothesis $Y_0=X$, and $Y_1=Y$ has the full strength of the alternative. We note that the scaling on the probability matrix $P$ results in a graph with moderate edge density and therefore moderate signal strength, and thus the hypothesis test we consider is a challenging one.

For a given matrix $X$, we generate $n_{\mathrm{boot}}=1000$ pairs of adjacency matrices from the RDPG model with latent positions $X$, compute the ASE for the two adjacency matrices, and get a bootstrapped sample test statistic value for this pair of estimated latent position matrices $\hat{X}_1,\hat{X}_2$. By choosing the 0.95 empirical quantile of the test statistics generated in this way for each of $T_F,T_S,T_R, \hat{T}_S,\hat{T}_R$, we obtain the bootstrapped critical value for each test statistic. Then for each alternative $Y_t$, $t\in(0,1]$, we again generate $n_{\mathrm{boot}}=1000$ pairs of adjacency matrices with $A\sim \mathrm{RDPG}(X), B\sim\mathrm{RDPG}(Y_t)$, compute their ASEs,
and compute the five test statistics on this pair $\hat{X},\hat{Y}_t$. Looking at the proportion of these computed test statistics exceeding the critical value for that test, we obtain a bootstrapped power for each strength of the alternative, allowing us to generate one power curve $p_i \colon [0,1] \to [0,1]$. By combining the $n_{\mathrm{MC}}=100$ power curves, we plot the mean power curve and error bars given by 
$$
\mathrm{mean}(t) \pm \frac{1.96}{\sqrt{n_{\mathrm{MC}}}}\mathrm{sd}(t),
$$
where the mean function $\mathrm{mean}(t)$ and standard deviation $\mathrm{sd}(t)$ are estimated from the $n_{\mathrm{MC}}$ samples $\{p_i(t)\}$ at each value of $t$.

From Figure~\ref{fig:diffuse-noise}, we see that in the case of diffuse noise, the Frobenius norm-based test statistic $T_F$ has the best power, followed by both Robust norm-based test statistics $T_R, \hat{T}_R$. The Spectral norm-based statistics $T_S, \hat{T}_S$ perform the most poorly, likely because they only identify a single dimension with the maximum variation, even though the deviation in this case is spread across all dimensions approximately equally. In this setting, the differences between $T_R$ and $\hat{T}_R$, as well as between $T_S$ and $\hat{T}_S$ are negligible as far as power is concerned, and the differences between these pairs of curves are nearly imperceptible.

\subsection{Rank-one alternative}


\begin{figure}
    \centering
\begin{tikzpicture}
  \begin{axis}[
    QuantilesLegend,
    xlabel={$t$}, ylabel={Power},
    width=.8\textwidth, height=.33\textheight,
    legend pos=north west,
    enlarge x limits=false,
    enlarge y limits=false
  ]
  \addplot[name path=upper1, draw, very thin,forget plot, TF!50] table[x=t, y=u, col sep=comma]{data/powercurves_csvs/fig2curve_F.csv};
  \addplot[name path=lower1, draw, very thin,forget plot, TF!50] table[x=t, y=l, col sep=comma]{data/powercurves_csvs/fig2curve_F.csv};
  \addplot[TF, forget plot, fill opacity=0.2] fill between[of=upper1 and lower1];
  \addplot[name path=upper2, draw, very thin,forget plot, TS!50] table[x=t, y=u, col sep=comma]{data/powercurves_csvs/fig2curve_SS.csv};
  \addplot[name path=lower2, draw, very thin,forget plot, TS!50] table[x=t, y=l, col sep=comma]{data/powercurves_csvs/fig2curve_SS.csv};
  \addplot[TS, forget plot, fill opacity=0.2] fill between[of=upper2 and lower2];
  \addplot[name path=upper3, draw, very thin,forget plot, TR!50] table[x=t, y=u, col sep=comma]{data/powercurves_csvs/fig2curve_RR.csv};
  \addplot[name path=lower3, draw, very thin,forget plot, TR!50] table[x=t, y=l, col sep=comma]{data/powercurves_csvs/fig2curve_RR.csv};
  \addplot[TR, forget plot, fill opacity=0.2] fill between[of=upper3 and lower3];
  \addplot[name path=upper4, draw, very thin,forget plot, TSF!50] table[x=t, y=u, col sep=comma]{data/powercurves_csvs/fig2curve_SF.csv};
  \addplot[name path=lower4, draw, very thin,forget plot, TSF!50] table[x=t, y=l, col sep=comma]{data/powercurves_csvs/fig2curve_SF.csv};
  \addplot[TSF, forget plot, fill opacity=0.2] fill between[of=upper4 and lower4];
  \addplot[name path=upper5, draw, very thin,forget plot, TRF!50] table[x=t, y=u, col sep=comma]{data/powercurves_csvs/fig2curve_RF.csv};
  \addplot[name path=lower5, draw, very thin,forget plot, TRF!50] table[x=t, y=l, col sep=comma]{data/powercurves_csvs/fig2curve_RF.csv};
  \addplot[TRF, forget plot, fill opacity=0.2] fill between[of=upper5 and lower5];
  \addplot[thick, TF] table[x=t, y=c, col sep=comma]{data/powercurves_csvs/fig2curve_F.csv};
  \addlegendentry{$T_{\mathrm{F}}$}
  \addplot[thick, TS] table[x=t, y=c, col sep=comma]{data/powercurves_csvs/fig2curve_SS.csv};
  \addlegendentry{$T_{\mathrm{S}}$}
  \addplot[thick, TR] table[x=t, y=c, col sep=comma]{data/powercurves_csvs/fig2curve_RR.csv};
  \addlegendentry{$T_{\mathrm{R}}$}
  \addplot[thick, TSF] table[x=t, y=c, col sep=comma]{data/powercurves_csvs/fig2curve_SF.csv};
  \addlegendentry{$\hat T_{\mathrm{S}}$}
  \addplot[thick, TRF] table[x=t, y=c, col sep=comma]{data/powercurves_csvs/fig2curve_RF.csv};
  \addlegendentry{$\hat T_{\mathrm{R}}$}
  \end{axis}
\end{tikzpicture}
    \caption{Power curves for rank-1 perturbation in span($X$), in dimension 3}
    \label{fig:rank1}
\end{figure}
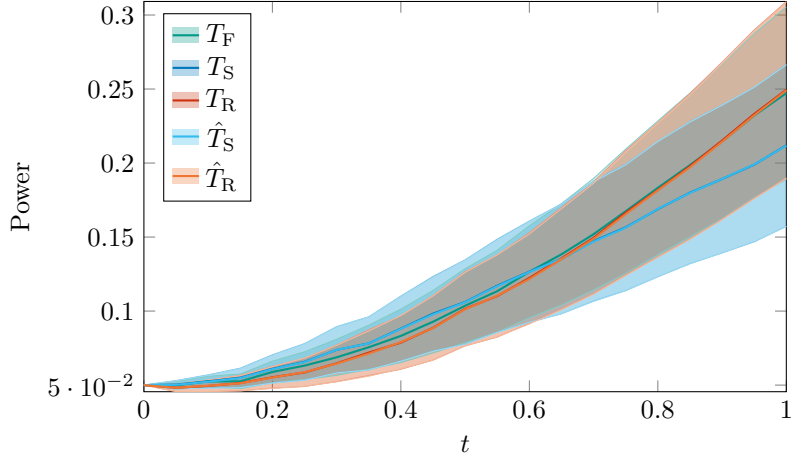


\begin{figure}
    \centering
\begin{tikzpicture}
  \begin{axis}[
    QuantilesLegend,
    xlabel={$t$}, ylabel={Power},
    width=.8\textwidth, height=.33\textheight,
    legend pos=north west,
    enlarge x limits=false,
    enlarge y limits=false
  ]
  \addplot[name path=upper1, draw, very thin,forget plot, TF!50] table[x=t, y=u, col sep=comma]{data/powercurves_csvs/fig3curve_F.csv};
  \addplot[name path=lower1, draw, very thin,forget plot, TF!50] table[x=t, y=l, col sep=comma]{data/powercurves_csvs/fig3curve_F.csv};
  \addplot[TF, forget plot, fill opacity=0.2] fill between[of=upper1 and lower1];
  \addplot[name path=upper2, draw, very thin,forget plot, TS!50] table[x=t, y=u, col sep=comma]{data/powercurves_csvs/fig3curve_SS.csv};
  \addplot[name path=lower2, draw, very thin,forget plot, TS!50] table[x=t, y=l, col sep=comma]{data/powercurves_csvs/fig3curve_SS.csv};
  \addplot[TS, forget plot, fill opacity=0.2] fill between[of=upper2 and lower2];
  \addplot[name path=upper3, draw, very thin,forget plot, TR!50] table[x=t, y=u, col sep=comma]{data/powercurves_csvs/fig3curve_RR.csv};
  \addplot[name path=lower3, draw, very thin,forget plot, TR!50] table[x=t, y=l, col sep=comma]{data/powercurves_csvs/fig3curve_RR.csv};
  \addplot[TR!10, forget plot, fill opacity=0.2] fill between[of=upper3 and lower3];
  \addplot[name path=upper4, draw, very thin,forget plot, TSF!50] table[x=t, y=u, col sep=comma]{data/powercurves_csvs/fig3curve_SF.csv};
  \addplot[name path=lower4, draw, very thin,forget plot, TSF!50] table[x=t, y=l, col sep=comma]{data/powercurves_csvs/fig3curve_SF.csv};
  \addplot[TSF, forget plot, fill opacity=0.2] fill between[of=upper4 and lower4];
  \addplot[name path=upper5, draw, very thin,forget plot, TRF!50] table[x=t, y=u, col sep=comma]{data/powercurves_csvs/fig3curve_RF.csv};
  \addplot[name path=lower5, draw, very thin,forget plot, TRF!50] table[x=t, y=l, col sep=comma]{data/powercurves_csvs/fig3curve_RF.csv};
  \addplot[TRF, forget plot, fill opacity=0.2] fill between[of=upper5 and lower5];
  \addplot[thick, TF] table[x=t, y=c, col sep=comma]{data/powercurves_csvs/fig3curve_F.csv};
  \addlegendentry{$T_{\mathrm{F}}$}
  \addplot[thick, TS] table[x=t, y=c, col sep=comma]{data/powercurves_csvs/fig3curve_SS.csv};
  \addlegendentry{$T_{\mathrm{S}}$}
  \addplot[thick, TR] table[x=t, y=c, col sep=comma]{data/powercurves_csvs/fig3curve_RR.csv};
  \addlegendentry{$T_{\mathrm{R}}$}
  \addplot[thick, TSF] table[x=t, y=c, col sep=comma]{data/powercurves_csvs/fig3curve_SF.csv};
  \addlegendentry{$\hat T_{\mathrm{S}}$}
  \addplot[thick, TRF] table[x=t, y=c, col sep=comma]{data/powercurves_csvs/fig3curve_RF.csv};
  \addlegendentry{$\hat T_{\mathrm{R}}$}
  \end{axis}
\end{tikzpicture}
    \caption{Power curves for rank-1 perturbation with angle $\pi/4$ to span($X$), in dimension 3}
    \label{fig:rank1-theta45}
\end{figure}
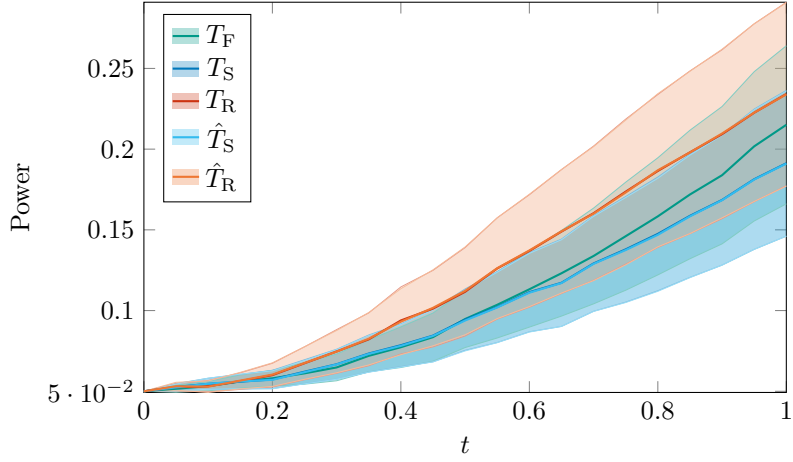


\begin{figure}
    \centering
\begin{tikzpicture}
  \begin{axis}[
    QuantilesLegend,
    xlabel={$t$}, ylabel={Power},
    width=.8\textwidth, height=.33\textheight,
    legend pos=north west,
    enlarge x limits=false,
    enlarge y limits=false
  ]
  \addplot[name path=upper1, draw, very thin,forget plot, TF!50] table[x=t, y=u, col sep=comma]{data/powercurves_csvs/fig4curve_F.csv};
  \addplot[name path=lower1, draw, very thin,forget plot, TF!50] table[x=t, y=l, col sep=comma]{data/powercurves_csvs/fig4curve_F.csv};
  \addplot[TF, forget plot, fill opacity=0.2] fill between[of=upper1 and lower1];
  \addplot[name path=upper2, draw, very thin,forget plot, TS!50] table[x=t, y=u, col sep=comma]{data/powercurves_csvs/fig4curve_SS.csv};
  \addplot[name path=lower2, draw, very thin,forget plot, TS!50] table[x=t, y=l, col sep=comma]{data/powercurves_csvs/fig4curve_SS.csv};
  \addplot[TS, forget plot, fill opacity=0.2] fill between[of=upper2 and lower2];
  \addplot[name path=upper3, draw, very thin,forget plot, TR!50] table[x=t, y=u, col sep=comma]{data/powercurves_csvs/fig4curve_RR.csv};
  \addplot[name path=lower3, draw, very thin,forget plot, TR!50] table[x=t, y=l, col sep=comma]{data/powercurves_csvs/fig4curve_RR.csv};
  \addplot[TR, forget plot, fill opacity=0.2] fill between[of=upper3 and lower3];
  \addplot[name path=upper4, draw, very thin,forget plot, TSF!50] table[x=t, y=u, col sep=comma]{data/powercurves_csvs/fig4curve_SF.csv};
  \addplot[name path=lower4, draw, very thin,forget plot, TSF!50] table[x=t, y=l, col sep=comma]{data/powercurves_csvs/fig4curve_SF.csv};
  \addplot[TSF, forget plot, fill opacity=0.2] fill between[of=upper4 and lower4];
  \addplot[name path=upper5, draw, very thin,forget plot, TRF!50] table[x=t, y=u, col sep=comma]{data/powercurves_csvs/fig4curve_RF.csv};
  \addplot[name path=lower5, draw, very thin,forget plot, TRF!50] table[x=t, y=l, col sep=comma]{data/powercurves_csvs/fig4curve_RF.csv};
  \addplot[TRF, forget plot, fill opacity=0.2] fill between[of=upper5 and lower5];
  \addplot[thick, TF] table[x=t, y=c, col sep=comma]{data/powercurves_csvs/fig4curve_F.csv};
  \addlegendentry{$T_{\mathrm{F}}$}
  \addplot[thick, TS] table[x=t, y=c, col sep=comma]{data/powercurves_csvs/fig4curve_SS.csv};
  \addlegendentry{$T_{\mathrm{S}}$}
  \addplot[thick, TR] table[x=t, y=c, col sep=comma]{data/powercurves_csvs/fig4curve_RR.csv};
  \addlegendentry{$T_{\mathrm{R}}$}
  \addplot[thick, TSF] table[x=t, y=c, col sep=comma]{data/powercurves_csvs/fig4curve_SF.csv};
  \addlegendentry{$\hat T_{\mathrm{S}}$}
  \addplot[thick, TRF] table[x=t, y=c, col sep=comma]{data/powercurves_csvs/fig4curve_RF.csv};
  \addlegendentry{$\hat T_{\mathrm{R}}$}
  \end{axis}
\end{tikzpicture}
    \caption{Power curves for rank-1 perturbation with angle $\pi/2$ to span($X$), in dimension 3}
    \label{fig:rank1-theta90}
\end{figure}
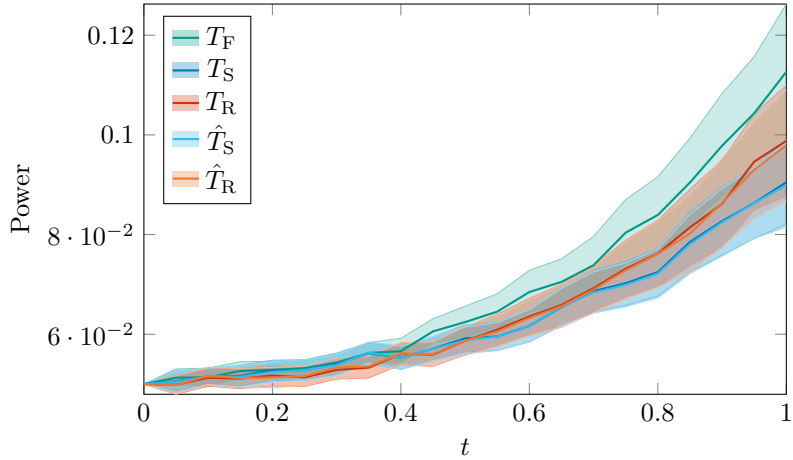

The procedure for generating the power curves in this case is similar to the previous one, but here the error takes the form $E=\sigma uv^{\top}$, where $u$ is chosen to lie (i) within $\mathrm{span}(X)$ ($\theta=0$), (ii) having an angle $\theta=\pi/4$ with this subspace, or (iii) being orthogonal to this subspace ($\theta=\pi/2$). For the first case, we generate $X$ as in the case of the diffuse alternative, then compute its QR factorization in order to obtain an orthonormal basis for its span. Next, $u$ is drawn as $Qz$ where $z\sim\mathcal{N}(0,I_r)$; $v$ is drawn according to $\mathcal{N}(0,I_r)$; and then $u$ and $v$ are normalized before forming $Y=X+uv^{\top}/2$. As before, to obtain matrices exhibiting different strengths of the alternative, we take $Y_t=(1-t)X+tY$, $t\in[0,1]$, and the rest of the estimation procedure is identical to the previous case.

To generate a rank-one perturbation having angle $\pi/4$ with $\mathrm{span}(X)$, we draw $a=Qz$, where $z\sim\mathcal{N}(0,I_r)$; then draw $b\sim \mathcal{N}(0,I_n)$, and take $c=(I-QQ^{\top})b$. After normalizing $a$ and $c$, we get $u=(a+c)/\sqrt{2}.$ For the fully orthogonal case, we draw $c$ in the same way, then set $u=c$.

The results in these settings, shown in Figures~\ref{fig:rank1}, \ref{fig:rank1-theta45}, and \ref{fig:rank1-theta90}, are more complicated than in the case of diffuse noise. First, the robust-norm based test statistic outperforms the others in the $\theta=0$ and $\theta=\pi/4$ cases. The Frobenius-norm based test statistic exhibits similar performance when $\theta=0$, but underperforms when $\theta=\pi/4$ and exceeds the robust-norm in the $\theta=\pi/2$ case. The $\theta=\pi/2$ case shows further separation between $T_R$ and $\hat{T}_R$ than any of the other settings considered, though even here the differences are relatively small. Across choices of $\theta$, the spectral-norm-based statistics $T_S$ and $\hat{T}_S$ exhibit comparable performance for low signal strengths, but lose steam compared to the robust and Frobenius-norm based test statistics in the presence of stronger alternatives.

\subsection{Salt-and-pepper alternative}

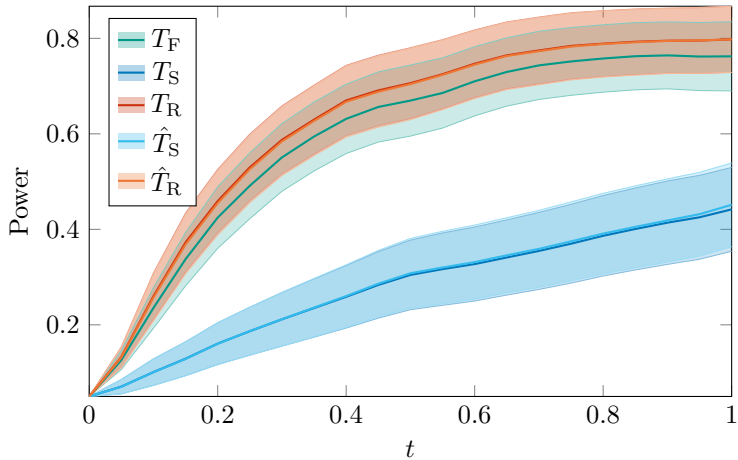
\begin{figure}
    \centering
\begin{tikzpicture}
  \begin{axis}[
    QuantilesLegend,
    xlabel={$t$}, ylabel={Power},
    width=.8\textwidth, height=.33\textheight,
    legend pos=north west,
    enlarge x limits=false,
    enlarge y limits=false
  ]
  \addplot[name path=upper1, draw, very thin,forget plot, TF!50] table[x=t, y=u, col sep=comma]{data/powercurves_csvs/fig5curve_F.csv};
  \addplot[name path=lower1, draw, very thin,forget plot, TF!50] table[x=t, y=l, col sep=comma]{data/powercurves_csvs/fig5curve_F.csv};
  \addplot[TF, forget plot, fill opacity=0.2] fill between[of=upper1 and lower1];
  \addplot[name path=upper2, draw, very thin,forget plot, TS!50] table[x=t, y=u, col sep=comma]{data/powercurves_csvs/fig5curve_SS.csv};
  \addplot[name path=lower2, draw, very thin,forget plot, TS!50] table[x=t, y=l, col sep=comma]{data/powercurves_csvs/fig5curve_SS.csv};
  \addplot[TS, forget plot, fill opacity=0.2] fill between[of=upper2 and lower2];
  \addplot[name path=upper3, draw, very thin,forget plot, TR!50] table[x=t, y=u, col sep=comma]{data/powercurves_csvs/fig5curve_RR.csv};
  \addplot[name path=lower3, draw, very thin,forget plot, TR!50] table[x=t, y=l, col sep=comma]{data/powercurves_csvs/fig5curve_RR.csv};
  \addplot[TR, forget plot, fill opacity=0.2] fill between[of=upper3 and lower3];
  \addplot[name path=upper4, draw, very thin,forget plot, TSF!50] table[x=t, y=u, col sep=comma]{data/powercurves_csvs/fig5curve_SF.csv};
  \addplot[name path=lower4, draw, very thin,forget plot, TSF!50] table[x=t, y=l, col sep=comma]{data/powercurves_csvs/fig5curve_SF.csv};
  \addplot[TSF, forget plot, fill opacity=0.2] fill between[of=upper4 and lower4];
  \addplot[name path=upper5, draw, very thin,forget plot, TRF!50] table[x=t, y=u, col sep=comma]{data/powercurves_csvs/fig5curve_RF.csv};
  \addplot[name path=lower5, draw, very thin,forget plot, TRF!50] table[x=t, y=l, col sep=comma]{data/powercurves_csvs/fig5curve_RF.csv};
  \addplot[TRF, forget plot, fill opacity=0.2] fill between[of=upper5 and lower5];
  \addplot[thick, TF] table[x=t, y=c, col sep=comma]{data/powercurves_csvs/fig5curve_F.csv};
  \addlegendentry{$T_{\mathrm{F}}$}
  \addplot[thick, TS] table[x=t, y=c, col sep=comma]{data/powercurves_csvs/fig5curve_SS.csv};
  \addlegendentry{$T_{\mathrm{S}}$}
  \addplot[thick, TR] table[x=t, y=c, col sep=comma]{data/powercurves_csvs/fig5curve_RR.csv};
  \addlegendentry{$T_{\mathrm{R}}$}
  \addplot[thick, TSF] table[x=t, y=c, col sep=comma]{data/powercurves_csvs/fig5curve_SF.csv};
  \addlegendentry{$\hat T_{\mathrm{S}}$}
  \addplot[thick, TRF] table[x=t, y=c, col sep=comma]{data/powercurves_csvs/fig5curve_RF.csv};
  \addlegendentry{$\hat T_{\mathrm{R}}$}
  \end{axis}
\end{tikzpicture}
    \caption{Salt-and-pepper noise in dimension 3}
    \label{fig:salt-and-pepper}
\end{figure}

In this case, we start by drawing matrices $X$ and $Y$ exactly as in the diffuse case, but the way that we obtain the matrices $Y_t, t>0$ is different from the previous cases. Here, we generate a permutation $\rho$ of $\{1,\ldots,n\}$, and for each value of $t\in (0,1]$, we let $n_t=\lfloor t n\rfloor$. The matrix $Y_t$ has rows equal to those of $X$ for indices that do not lie in the set $\{\rho(1),\ldots,\rho(n_t)\}$, and has rows equal to those of $Y$ for indices that do lie in this set. Thus, as $t$ increases, an increasing proportion of the rows of $X$ are replaced with ones that are distributionally similar, but not equal. Besides this change, the rest of the estimation procedure is identical.

The results shown in Figure~\ref{fig:salt-and-pepper} demonstrate a clear preference for the robust-norm-based test statistics, followed by the Frobenius norm. Meanwhile, the spectral-norm-based test statistics dramatically underperform. We note that in this setting, the signal strength is considerably stronger for $t=1$ than in the previous settings, owing to the method by which we generate the pairs $(X,Y)$. However, even for small values of $t$, we see a very rapid separation between the top performers and the spectral-norm-based statistics. Once again, there is a minimal separation between the test statistics $T_R/\hat{T}_R$ and $T_S/\hat{T}_S$ in this case, indicating that the Frobenius-norm solution does not noticeably impact the power of these test statistics.

\section{Discussion}\label{sec:discussion}

Motivated by challenges in both optimization and statistics, we consider Procrustes problems for different choices of matrix norm: Frobenius, Spectral, and Robust. Recent results in Riemannian optimization provide algorithms for finding local solutions to Procrustes problems with norms other than the Frobenius norm, and we deploy these new algorithms to solve a wider array of Procrustes problems that are important for statistical network inference. Our simulation results on hypothesis testing for networks shed light on when and how the solutions to these Procrustes minimizations differ, and what impact this has on statistical power. We find the surprising result that in several instances, approximately solving the Spectral or Robust Procrustes minimization using the easily-computed Frobenius-norm minimizer does not significantly degrade inference. This gives ballast to its use as a computationally inexpensive shortcut. We also examine how these different alignment problems are impacted by types of perturbations of matrix entries. We find that the Robust Procrustes test statistic performs well under a variety of alternative hypotheses for network hypothesis testing, but can perform poorly under contamination by certain global errors (see Figure~\ref{fig:duck:result-twospec}). 

Given that we have only considered local solvers of the Spectral Procrustes problem without a global analysis of the problem, it may be possible to obtain a more detailed characterization of the solution to this problem from a linear-algebraic perspective. On the other hand, the performance of the robust norm under various alternatives suggests using this as a test statistic for network hypothesis testing and conducting a theoretical power analysis of this and other test statistics under various models. As a practical matter, with these new optimization methods, statisticians can now compute solutions to Procrustes problems under alternative choices of norm. 

Future research directions include the analysis and development of solvers that can solve nonsmooth generalized Procrustes problems with provably certified global optimality under appropriate conditions, which is beyond the scope of this study. The design of solvers based on low-rank semi-definite programmaing (SDP) reformulations \cite{fulconic}, so far rather unexplored, might play a role in this. Scalability of the solvers to high-dimensional latent position vectors $d \gg 1$ and for large networks $n \gg 1$ is an important aspect of any solver to be considered for real-world problem instances. Our simulations have shown that Procrustes problems with robust objectives may outperform spectral objectives for certain statistical applications. Going beyond norms and convex penalizations \eqref{eq:RobustProcrustes} and considering a non-convex penalization, as often done in robust statistics and computer vision \cite{Black1996unification,Mactavish2015all,  Yang-2019Quaternion,Peng2023convergence,Huang2024Scalable} is a promising direction to obtain alignment methods that are even more robust to outliers than the methodologies considered in this paper.

Understanding the conditions under which the solutions to these problems differ is a key next step in theoretical developments, and connects to ongoing research on numerical and statistical error, the choice of appropriate matrix norm for a given inference task, the optimization methods required for different norms, and the impact of computational shortcuts on accuracy in estimation and testing. This reinforces the relevance of further work on Procrustes problems in optimization and statistics.

\appendix

\printbibliography

\end{document}